# Properties of the Nili Fossae Olivine-clay-carbonate lithology: orbital and in situ at Séítah

Adrian J. Brown*[1], Linda Kah[2], Lucia Mandon[3], Roger Wiens[4], Patrick Pinet[5], Elise Clavé[6], Stéphane Le Mouélic[7], Arya Udry[8], Patrick J. Gasda[9], Clément Royer[10], Keyron Hickman-Lewis[11], Agnes Cousin[12], Justin I. Simon[13], Jade Comellas[14], Edward Cloutis[15], Thierry Fouchet[16], Alberto G. Fairén[17], Stephanie Connell[15], David Flannery[18], Briony Horgan[4], Lisa Mayhew[19], Allan Treiman[20], Jorge I. Núñez[21], Brittan Wogsland[2], Karim Benzerara[22], Hans E.F. Amundsen[23], Cathy Quantin-Nataf[25], Kevin P. Hand[26], Vinciane Debaille[27], Ari Essunfeld[9], Pierre Beck[28], Nicholas J. Tosca[29], Juan M. Madariaga[30] and Eleni Ravanis[31]

[1]Plancius Research, Severna Park, MD 21146. [2]Department of Earth and Planetary Sciences, University of Tennessee, Knoxville TN 37996, [3]LESIA, Observatoire de Paris, Université PSL, CNRS, Sorbonne Université, Université de Paris, Meudon, France, [4]Earth, Atmospheric, and Planetary Sciences, Purdue University, [5]IRAP, Observatoire Midi-Pyrénées,CNRS, Toulouse university, France, [6]CELIA, Université de Bordeaux, CNRS, CEA, [7]Laboratoire Planétologie et Géosciences, CNRS UMR 6112, Nantes Université, Université Angers, 44322 Nantes, France, [8]Department of Geoscience, University of Nevada Las Vegas, 89154 Las Vegas, NV, [9]Los Alamos National Laboratory, PO Box 1663 Los Alamos, NM 87545, [10]LESIA, Observatoire de Paris, Université PSL, CNRS, Sorbonne Université, Université de Paris, Meudon, France, [11]The Natural History Museum, Cromwell Road, London,  SW7 5BD, UK, [12]IRAP, Observatoire Midi-Pyrénées, CNRS, Toulouse university, France, [13]Center for Isotope Cosmochemistry and Geochronology, Astromaterials Research & Exploration Science Division, NASA Johnson Space Center, 2101 Nasa Parkway, Houston, TX, 77058 USA, [14]University of Hawai'i at Mānoa, [15]University of Winnipeg, Canada, [16]LESIA, Observatoire de Paris, Université PSL, CNRS, Sorbonne Université, Université de Paris, Meudon, France, [17]Centro de Astrobiologia (CSIC-INTA) & Cornell University, NY, [18]Queensland University of Technology, Queensland, Australia, [19]Department of Geological Sciences, University of Colorado Boulder, Boulder, CO 80309, [20]Lunar and Planetary Institute, Houston TX 77058, [21]Johns Hopkins University Applied Physics Laboratory, Laurel, MD 20723, [22]Institut de Minéralogie, Physique des Matériaux et Cosmochimie, CNRS, Museum National d'Histoire Naturelle, Sorbonne Université, Paris, France, [23]CENSSS, University of Oslo, Norway, [25]University of Lyon, France, [26]Jet Propulsion Laboratory, Pasadena, CA 91109, [27]Laboratoire G-Time, Université libre de Bruxelles, Brussels, Belgium, [28]Institut de Planétologie et d'Astrophysique de Grenoble IPAG/CNRS, Université Grenoble Alpes, 414 rue de la piscine, 38041 Grenoble cedex 09, France, [29]Department of Earth Sciences, University of Cambridge, Downing Street, Cambridge, CB2 3EQ, United Kingdom, [30]Department of Analytical Chemistry, University of the Basque Country UPV/EHU, 48940 Leioa, Spain, [31]Hawai'i Institute of Geophysics and Planetology, University of Hawai'i at Mānoa, Honolulu, HI, USA

Corresponding author: Adrian J. Brown (adrian.j.brown@nasa.gov)

**Key Points:**

- **We use a Venn diagram approach to show the Nili Fossae olivine lithology is better named olivine-clay-carbonate**
- **We postulate a flood lava origin for the olivine-clay-carbonate lithology at Séítah based on finding olivine cumulate and low viscosity lava**
- **We find the clay in the cumulate olivine at Séítah is either talc or serpentine and eliminate other clay families**

**Plain Language Summary**

We used orbital and in situ data to observe a lava flow near the Mars 2020 landing site at Jezero Crater. By analyzing the reflectance spectra of the rocks containing the lava, we have identified that clay is present in the rocks. We use in situ imaging data to determine that the lava contains close packed crystals (cumulate), a process which can happen in the bottom of a lake of lava. We use measurements from the SuperCam LIBS instrument to determine that the cumulate is accompanied by clays and eliminate families of clays to conclude it is either talc or serpentine.

**Abstract**

We examine the observed properties of the Nili Fossae olivine-clay-carbonate lithology from orbital data and in situ by the Mars 2020 rover at the Séítah unit in Jezero crater, including: 1) composition (Liu, 2022) 2) grain size (Tice, 2022) 3) inferred viscosity (calculated based on geochemistry collected by SuperCam (Wiens, 2022)). Based on the low viscosity and distribution of the unit we postulate a flood lava origin for the olivine-clay-carbonate at Séítah.

We include a new CRISM map of the clay 2.38 μm band and use in situ data to show that the clay in the olivine cumulate in the Séítah formation is consistent with talc or serpentine from Mars 2020 SuperCam LIBS and VISIR and MastCam-Z observations.

We discuss two intertwining aspects of the history of the lithology: 1) the emplacement and properties of the cumulate layer within a lava lake, based on terrestrial analogs in the Pilbara, Western Australia, and using previously published models of flood lavas and lava lakes, and 2) the limited extent of post emplacement alteration, including clay and carbonate alteration (Clave, 2022; Mandon, 2022).

# 1. Introduction

*1.1 Purpose of paper*

We examine the observed properties of the Nili Fossae olivine-clay-carbonate lithology, including: 1) composition (Liu, 2022) and (Corpolongo, 2022), 2) grain size (Tice, 2022) and size frequency distribution (Kah, 2022), and 3) inferred viscosity (calculated based on geochemistry of (Wiens, 2022)).

*Importance of Viscosity.* Why are we interested in the viscosity of the olivine-clay-carbonate lithology? As we shall discuss further in this paper, the viscosity is a key characteristic of a lava flow, determined by a variety of factors, including its composition, temperature, water and crystal content and pressure. The viscosity is a critical parameter for controlling the morphologic expression of the erupted unit. Critically, it allows us to compare the lava with other similar flows across Mars, and place it in context with the dominant rock-forming process on the planet. The inferred viscosity has now become available from the first in situ measurements of Jezero crater floor geochemistry.

Previously the lithology has been known as olivine-bearing (Hoefen et al., 2003; Tornabene et al., 2008) or olivine-carbonate (Ehlmann et al., 2008; Mandon et al., 2020; Brown et al., 2020), however, this paper is designed to quantitatively analyze the relationship between olivine, carbonate and clay and investigate, using a Venn diagrammatic approach; the suggestion being that where there is carbonate, there is also clay (Horgan et al., 2020; Tarnas et al., 2021). Here we present a new CRISM map of the Fe-Mg clay 2.38 μm band and show major and minor elements using in situ observations from SuperCam Laser Induced Breakdown Spectroscopy (LIBS) and Visible and Shortwave Infrared (VISIR). We conclude that the clay in the Séítah unit is most consistent with talc over serpentine.

We address in detail two aspects of the olivine-clay-carbonate lithologic history: 1.) the emplacement of the cumulate layer within a lava lake, which is part of a flood lava, based on terrestrial analogs in the Pilbara, Western Australia, and previously published models of flood lava eruption and flow, and lava ponding in lakes (Worster et al., 1993), and 2.) the limited extent of post emplacement alteration, including clay and carbonate alteration of the Séítah formation (Clave, 2022; Mandon, 2022). These possibly quite disparate aspects of the history of the unit shall drive this paper, which is designed to show how observations and models are both needed to discover further insights into the emplacement and subsequent alteration history of this important Martian lithology, which has now been sampled for return to Earth (Simon, 2022).

*Unit Interpretations.* Farley et al. (2022) list definitions for the rock lithologies we will discuss in this paper. Farley et al. defined the Séítah (an olivine cumulate) and Máaz (pyroxene- and plagioclase-dominated overlying Séítah) formations that were encountered and mapped during the Crater Floor campaign in the first 380 sols of the Mars 2020 mission. In this paper we shall discuss interpretations arising from the Séítah formation. Séítah has previously been interpreted as related to the olivine-carbonate lithology inside and outside Jezero, and our study shall make that assumption too at this stage, although it is yet to be definitively proven since the *Perseverance* rover has not encountered the olivine-carbonate lithology in the marginal carbonates near the crater rim, or outside Jezero. Figure 1 presents a visual summary of the traverse through the Séítah formation and the three key workspaces, and all the targets referred to in this paper. On Sol 202 of the mission the rover rolled into the Séítah formation, and approached the Bastide workspace. For the purposes of this study, the rover visited three main workspaces - in temporal order, Bastide, Brac and Issole. These workspaces and the abrasions and samples taken at each workspace are also shown in Figure 1. The rover left Séítah on Sol 340. See also (Sun, 2022; Crumpler, 2022; Núñez, 2022) for more details on the Crater Floor campaign and stratigraphic information obtained during the Séítah traverse.

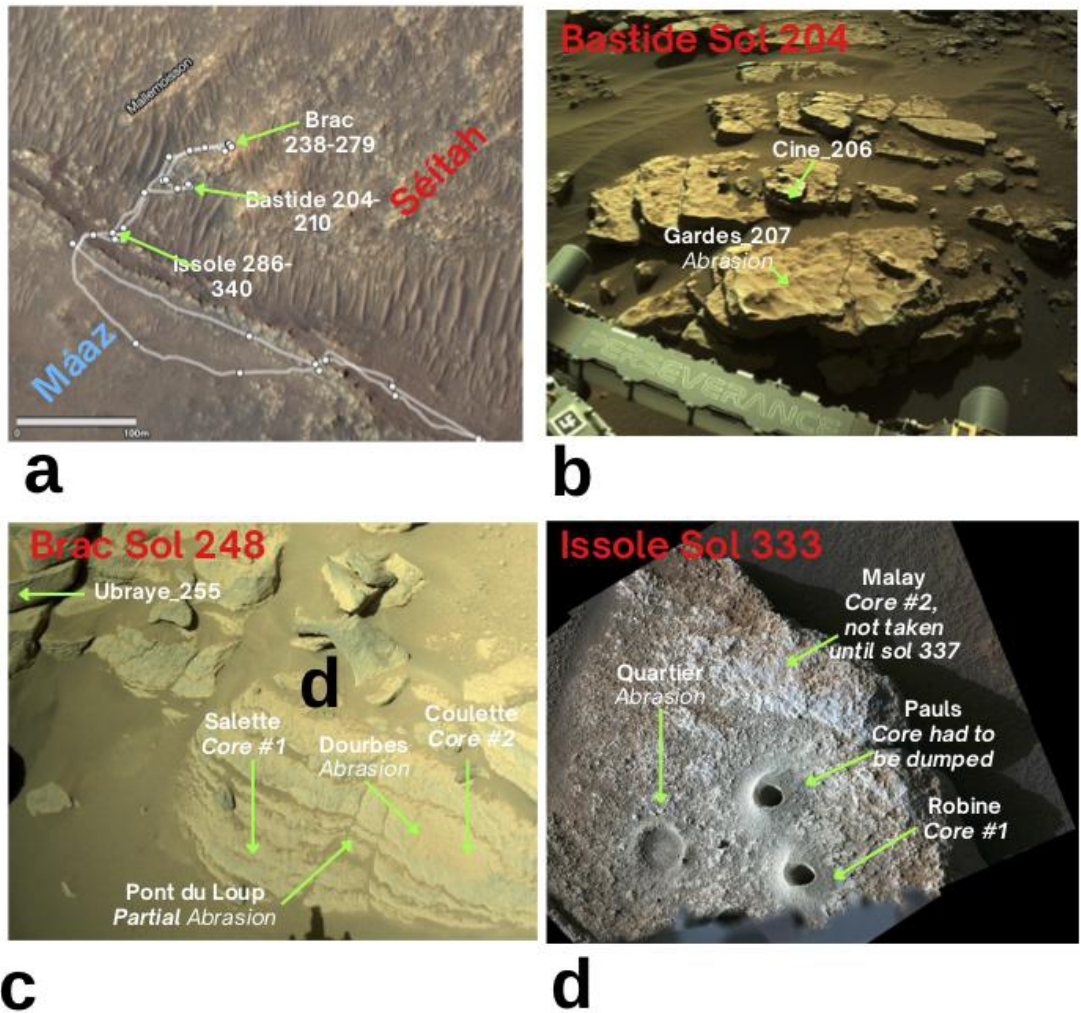

Figure 1. a.) HiRISE overview map of the Séítah traverse showing the three key workspaces: Bastide, Brac, and Issole, and the Sol counts at each. b.) Navcam image of Bastide Workspace taken on Sol 204. c.) Navcam image of Brac Workspace on Sol 248, before samples were taken NLF_0248_0688961170_442ECM_N0080000NCAM02248_07_195J d.) Mastcam-Z image of Issole on Sol 333 after Robine and Pauls samples and Quartier abrasion were taken but before sample Malay was obtained, QZCAM_SOL0333_ZCAM08355_L0_Z110_ISSOLE_WORKSPACE_VP_4X_E01.

*Definition of clay*. Due to its importance in this paper, we first establish a clear and distinct definition of the word "clay". Here by "clay" we are referring to the mineral group of phyllo-silicates ("φύλλο" in Greek is "leaf" in English). These layer-silicates have stacked hydroxyl (-OH), octahedral (O) sheets and tetrahedral (T) silicate (Bailey, 1988). The octahedral sheets come in two forms called dioctahedral and trioctahedral which host divalent ($Fe^{2+}/Mg^{2+}$) and trivalent (e.g., $Al^{3+}/Fe^{3+}$) cations respectively. Clays come in many families, based on the stacking of layers (e.g. TO or 1:1 vs TOTTOT or 2:1). In this paper we are particularly interested in the 2:1 phyllosilicates talc and 1:1 serpentine and the 2:1 clay family called smectites. We shall discuss these more below.

*1.2 Previous work*

*1.2.1 Infrared remote observations*

The Syrtis Major shield region was first observed to host olivine using Visible and Near InfraRed (VNIR) telescopic observations by Pinet and Chevrel (1990). Spacecraft observations followed, and a decade later the Thermal Emission Spectrometer (TES) infrared (IR) instrument also recognised a strong 10 μm absorption band associated with the Nili Fossae region that was attributed to olivine (Hoefen et al., 2003). The Observatoire pour la Minéralogie, l'Eau, le Glace et l' Activité (OMEGA) instrument on Mars Express was able to determine that phyllosilicates/clays were also associated with the olivine, suggesting Fe-rich smectites were present (Mangold et al., 2007). Using data from the Compact Reconnaissance Imaging Spectrometer for Mars (CRISM) instrument, Mg-carbonate was discovered to be associated with the olivine lithology by using the strength and position of the 2.5 μm band (Ehlmann et al., 2008). The olivine-carbonate was hypothesized to be due to serpentinization by Brown et al. (2010), who also suggested that talc might be present by analogy to talc-carbonate terrestrial Archean greenstone terranes. Brown et al. (2010) also found the carbonate to be a 50/50 Fe/Mg mixture, which was later supported using TES orbital data analyzed by Ruff et al. (2022). Goudge et al. (2015) mapped the watershed of the Jezero deltas and the accompanying mineralogy. Mafic minerals and marginal carbonates were identified in the mapping of Horgan et al. (2020). Mandon et al. (2020) used OMEGA observations to map the olivine and carbonate, and used crater counting to date the unit at 3.82 Ga. The olivine composition (Fo#44-66) and grain size (~ 1mm) were calculated by Brown et al. (2020) using CRISM data .

*1.2.2 SuperCam LIBS observations*

SuperCam is a multi-technique instrument suite mounted on the Mars 2020 *Perseverance* rover (Wiens et al., 2021; Maurice et al., 2021). It is an upgraded version of the ChemCam instrument onboard the Mars Science Laboratory (MSL) *Curiosity* rover. It has the capability to obtain LIBS measurements to a distance of ~10 m, and to quantify the major element abundances at distances up to 6.5 m. The Remote Micro-Imager (RMI) is a color telescope used to provide context images for SuperCam observations.

The SuperCam instrument has been used to investigate the igneous nature of the Séítah and Máaz regions, as reported by Wiens et al. (2022). That study concentrated on chemistry and mineralogy, and creation of a stratigraphy of the two regions of the Jezero crater floor. In this paper we use LIBS elemental abundances for the Séítah and Máaz formations reported in the Wiens et al. (2022) study.

*1.2.3 SuperCam VISIR observations*

SuperCam VISIR is an optical spectrometer that covers the 0.4-0.85 and 1.3-2.6 µm wavelength ranges, providing reflectance spectra boresighted with LIBS and RMI instruments (Fouchet et al., 2022). SuperCam VISIR measurements are reported in two accompanying papers in this special collection (Royer, 2022; Mandon, 2022). These papers introduce the calibration of the VISIR instrument using laboratory analysis and measurements of calibration targets on Mars (Royer, 2022) and all the previous SuperCam VISIR results taken on the Martian surface (Mandon, 2022). This paper relies on the calibration and comparison of the spectra presented here with others measured on the surface. Royer et al. (2022) shows that, based on band depth calculations and noise estimates, that most of the 2.46 µm detections we will discuss herein range from Signal to Noise Ratio (SNR) = 4 to SNR = 12. This spectral range is harder to study because of the fact that many spectral signatures are sampled with the minimum sampling (at most 2 spectral channels in the band) and several artifacts may be similar to absorption features (e.g., non-corrected spikes, calibration residuals). In our case, regarding the 2.46 µm region, we have SNR > 5 and as demonstrated below, detections of small bands in this region are realistic, given the accuracy of the calibration and the potential residuals. Positive detections must nevertheless be confirmed visually and their attribution validated by the presence of other features characteristic of phyllosilicates.

SuperCam's VISIR data allow on-ground comparison with absorption features seen from orbit by the CRISM instrument. Because CRISM covers a larger spectral range (0.4-3.96 µm), in this paper we adopt a band fitting strategy over the 1.3 to 2.5 µm range. We use an asymmetric Gaussian shape to fit the data and directly compare the band shape, asymmetry, halfwidth and amplitude.

*1.2.4 PIXL observations*

PIXL is a high resolution X-Ray fluorescence microscope in situ scanning instrument mounted on the arm turret of the rover (Allwood et al., 2020). PIXL observations of the Dourbes target have been used to find the cumulate nature of the target and infer an igneous origin (Liu, 2022). They have been used to infer that the olivine composition is consistently homogeneous, and of Fo55+/-1 within a several mm analysis patch (Liu, 2022), which is a remarkable match with previous orbital work constraining the olivine Fo# to 44-66 (Brown et al., 2020). PIXL's X-ray diffraction capability, which can identify coherent crystalline domains, has been used to identify coarsely grained (1-2 mm), interlocking olivine grains (Tice, 2022), and SuperCam's RMI instrument measured a mean grain size of 1.45+/-0.20 mm over the rover's Seitah formation traverse (Wiens et al. 2022). These in situ observations are both good matches for the ~1 mm Nili Fossae olivine grain size calculation from orbital studies prior to landing (Brown et al., 2020).

## 2. Methods

*2.1 Viscosity calculations*

*2.1.1 Bottinga and Weill empirical melt viscosity model*

In order to calculate the predicted melt viscosity, we have used the empirical approach of Bottinga and Weill (1972), which calculates a viscosity vs. temperature curve for the magmatic silicate liquid. They present five tables according to the silica content of the magma. They then use these tables to sum the contributions of the elements common to terrestrial magma bodies to arrive at a predicted viscosity for bodies of that composition.

*2.1.2 Flood lava and lava lake models*

Martian gravity is ~37.5 % of that on Earth. Assuming Mars possessed a thinner (~6 mbar) atmosphere throughout its history, we can compare terrestrial lava properties to those of Mars (Wilson and Head, 1994). On Mars, with a thinner atmosphere, lava will take longer to cool. Longer cooling rates should also drive larger crystal growth. Wilson and Head (1994) predicted Martian volcanic flows (driven by lower gravity) to be 6 times as long as compositionally identical flows on Earth, leading to spatially larger deposits on the smaller planet.

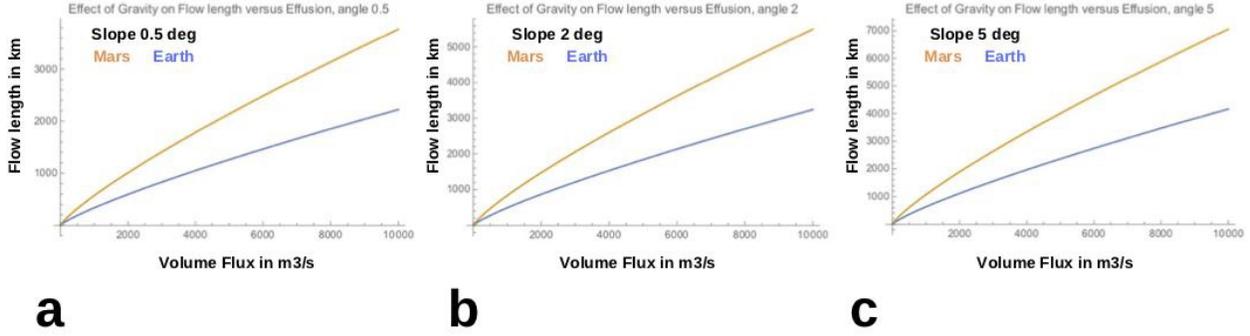

Figure 2. Lava flow lengths for Martian and Earth conditions for three different ground slopes. Modeled after Wilson and Head (1994), and recalculated using eqn (1).

Pinkerton and Wilson (1994) presented a Bingham non-isothermal model for calculation of the runout lengths for lavas on Earth. This model depends upon viscosity, so we used this equation to calculate the expected runout length $L$ in Figure 2 for Mars and Earth gravity. This is Equation 7 of Wilson and Head (1994):

$$L = [1.34/\kappa G_{zc}](\tau/\eta)^{2/11} E^{9/11} [\tau/(\rho g)]^{6/11} \sin^{3/11} \alpha \tag{1}$$

where $\kappa$ is the thermal diffusivity of lava, taken as $7 \times 10^{-7}$ m$^2$/s, $G_{zc}$ is the critical Grätz number, taken as 300, $\tau$ is the yield strength of the lava, taken as 2000, $\eta$ is the viscosity, $(\tau/\eta)^{2/11}$ is taken as 0.38 as suggested by Wilson and Head (1994), $\rho$ is the density of the lava, taken here as a value appropriate for ultramafic lavas of 3200 kg/m$^3$, g is the acceleration due to gravity, 9.81m/s$^2$ for Earth and 3.72m/s$^2$ for Mars, $\alpha$ is the average slope angle in degrees of the ground over which the lava flows. The results of this calculation are shown in Figure 2.

The Reynolds number of a liquid flow can be calculated (Huppert et al., 1984):

$$R_e = E/\eta \tag{2}$$

Here $E$ is the discharge rate and $\eta$ is the kinematic viscosity. In essence, $R_e$ is a measure of the "friskiness" of the lava - if the viscosity is low and magma is rapidly erupting, it will be high. The lava will flow readily and cover significant terrain.

At high Reynolds number, the discharge rate $E$ of lava through a long fissure of width $d$ can be calculated thus:

$$E = d^{3/2}(g\Delta\rho)^{1/2}/(k\rho_m)^{1/2} \tag{3}$$

Here $k$ is a friction coefficient, $\Delta\rho$ is the density difference between the magma and lithosphere, and $\rho_m$ is the magma density.

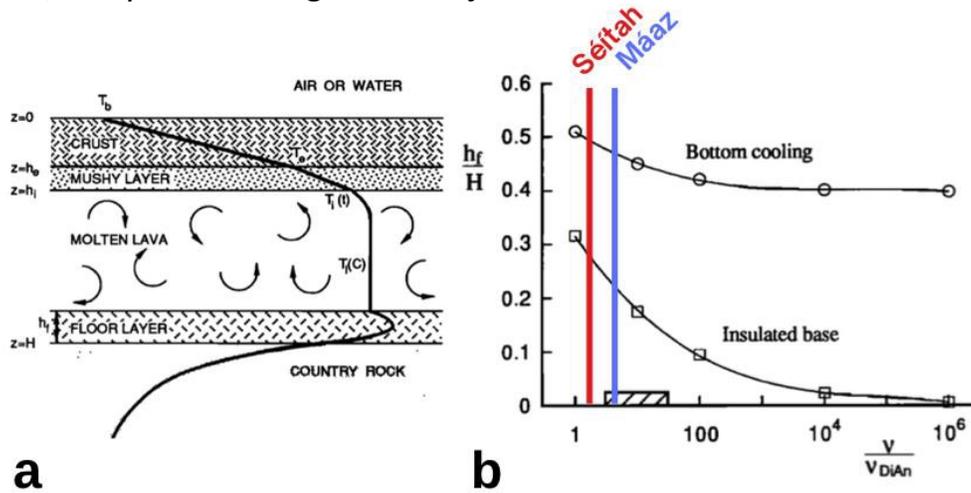

Figure 3. a.) Thermal model of a lava flow with layers representing solid crust/cap, mushy layer (where crystals form), molten lava (with convection taking place) and the floor. b.) Final fractional height of floor layer, $h_f$, where crystals accumulate (relative to the height of the chamber height, $H$) plotted against the lava viscosity (normalized relative to liquid diopside). Two cases are shown, one in which bottom cooling occurs, and one in which it does not (insulated base). The dashed box indicates the viscosity of typical terrestrial basalt relative to a pure diopside melt, after Worster et al. (1993). The Séítah and Máaz viscosities calculated in this paper are schematically shown on the plot.

*Development of layering*. The large grain size (Kah, 2022) and cumulate texture (Liu, 2022; Tice, 2022) lead us to explore the possibility of the Séítah rocks being emplaced as a martian lava lake. Worster et al. (1993) developed a thermodynamic model for a terrestrial lava lake, and outlined how the body would form compositional and textural layers within the resultant rock as it cooled. They showed that a lava lake can develop quickly when a magma body is exposed to the atmosphere, and showed that convective cooling can take place. They examined the situation with and without convection in their model and found thermal convection drives internal crystallization within the melt. This in turn forms layering within the floor of the lake. Thermal conduction alone does not produce this layering. Figure 3a shows a schematic diagram of an idealized lava lake after Worster et al. (1993). For our purposes, we will define a lava lake as a connected igneous body that is cooling convectively and develops internal stratification (i.e. layering) as shown in this diagram.

*Effect of viscosity*. Figure 3b shows the effect of viscosity on the cooling history of a lava lake that is cooled through the bottom. This figure shows the major effect on lava lake crystallization studied by Worster et al. (1993). The parameter $h_f$ is the final floor layer (Figure 3a) depth, and $H$ is the total depth of the lava lake. It can be seen that the relative size of the floor layer grows with decreasing viscosity, and this effect is more pronounced when the bottom of the rock is insulated. We show schematically with vertical lines where the Séítah and Máaz units would plot in the viscosity (relative to liquid diopside), indicating that Séítah-like viscosities would produce more crystallization than a Máaz-like viscosity.

The takeaway message from this section is that lower viscosity lavas build deeper cumulate layers. We shall use this model to enhance the interpretations of our in situ observations later in this study.

## 2.2 Clay alteration mineralogy

### 2.2.1 CRISM and VISIR clay identification

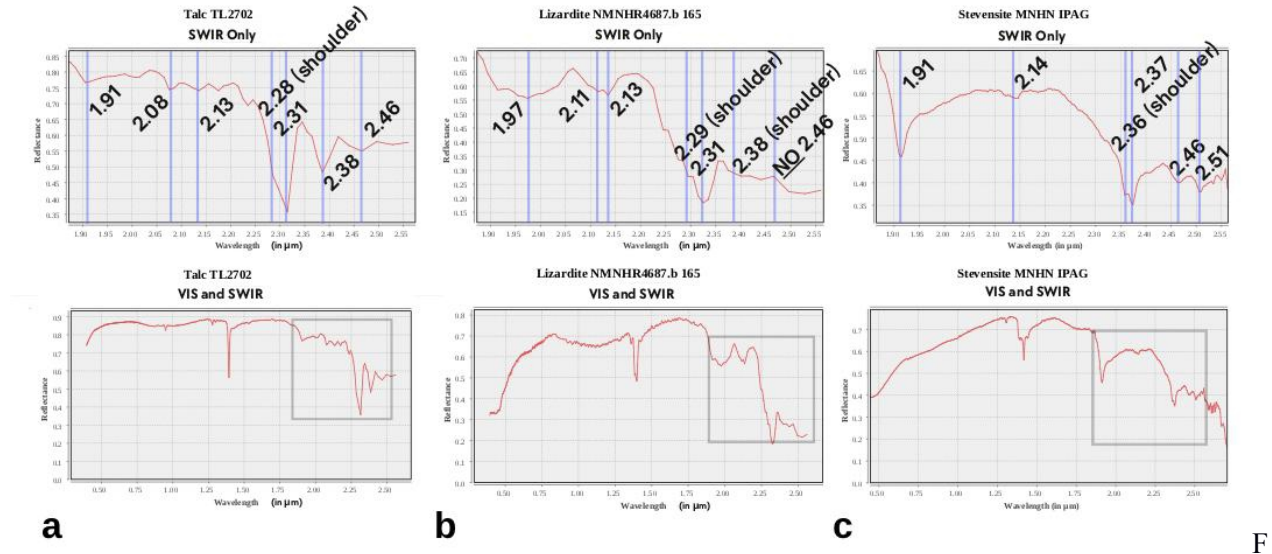

Figure 4. Laboratory spectra of a.) Talc b.) Lizardite c.) Stevensite. Shown at bottom is the visible and SWIR region of the spectrum, on top is a zoom into the SWIR region at the location shown in the box on the lower spectra. Note: 1) strength of 2.31 μm band in talc and 2) the presence of a weak 2.46 μm band in talc which is not present in lizardite. Talc and lizardite spectra are from USGS Spec Library (Clark et al., 2007), stevensite sample is from the Muséum National d'Histoire Naturelle collection provided to P. Beck.

Figure 4 shows library spectra of talc, the serpentine lizardite, and stevensite, in the VIS and SWIR regions, which are covered by the SuperCam VISIR instrument. The stevensite specimen was provided by the Museum National d'Histoire Naturelle in Paris. The sample was analyzed using the SHADOWS instrument (Potin et al., 2018) in micro-beam mode (2 mm).

It can be seen from Figure 4c that this example of stevensite has a very different SWIR spectrum to the talc and lizardite (e.g. the shift of the 2.31 μm band to 2.36 μm in stevensite), so we do not discuss it further in our comparisons.

We also note that the SWIR absorption bands of the Fe-bearing clays nontronite (Bishop et al., 2013) and hisingerite (Tosca, 2022; Turrenne et al., 2022) both display strong 2.28 μm bands which are not present in the spectra of olivine bearing targets we report here, so we do not discuss them further.

The key bands of the talc and lizardite library spectra are noted in the plot, and we draw the reader's attention to the relative strength of the 2.31 (strong), 2.38 (moderate to weak) and 2.46 μm (weak) bands in particular. The two key differences for the purpose of this paper are 1) the strength of the 2.38 μm band, which is present in serpentine, but far stronger in talc, and 2) the presence of a weak 2.46 μm band in talc which is only a shoulder (at most) in lizardite. These two key spectral differences are the focus of the remainder of the paper.

*2.2.1.2 Band fitting with an Asymmetric Gaussian shape*

As has been previously discussed in a variety of forums such as Brown (2006), the natural bandshape for an absorption band in energy space is a Gaussian shape. When accompanied by nearby and often closely related bands, for example, due to Tschermak substitution of $Mg^{2+}$ and $Fe^{2+}$ (Duke, 1994), the vibration bands can adopt an asymmetric shape. It is then advantageous to carry out an asymmetric band fit. Figure 4 demonstrates that all three of the clay related bands are asymmetric and extend to the left (higher energy). As part of this fitting process we obtain four parameters per band: the amplitude, width, centroid (position), and the asymmetry (Brown et al., 2010). We use these band parameters in our study below.

*2.2.2 Chemical formulas of serpentine, talc and smectites*

The ideal chemical formula of lizardite (Mg-serpentine) is $Mg_3Si_2O_5(OH)_4$, and that of Mg-talc is $Mg_3Si_4O_{10}(OH)_2$. Kerolite is a hydrated, poorly crystalline talc-like phase ($Mg_3Si_4O_{10}(OH)_2 \cdot nH_2O$) (Brindley et al., 1977). For the purposes of this study, we point out that there is (ideally) no aluminum in their structure. This is in contrast to smectites, which have a complex interplay of substitution, even from crystal to crystal in a sample (García-Romero et al., 2021). For our purposes, this smectite family includes four endmembers: 1) saponite (Treiman et al., 2014;

Roush et al., 2015) the formula of which can vary but may have the following value: $Ca_{0.1}Na_{0.1}Mg_{2.25}Fe^{2+}_{0.75}Si_3Al_1O_{10}(OH)_2 \cdot 4(H_2O)$, 2) beidellite, 3) stevensite, and 4) montmorillonite (Bishop et al., 2008). All of these smectites (except stevensite, (Tosca and Wright, 2018)) have Al in their crystal structures. This difference in aluminum content will be used in the following section to discriminate between the candidate clay minerals.

## 3. Results

### 3.1 Viscosity results and lava lake calculations

Table 1 presents the elemental abundances derived using SuperCam LIBS measurements on rocks from the Máaz formation, Artuby member and Séítah formation (Wiens, 2022). The bottom line of the table shows the calculated viscosity at 1470 K. This calculation was conducted using the Bottinga-Weill empirical melt viscosity technique described earlier. The results of this calculation are plotted against the inverse temperature in Figure 5a. We have presented their viscosity results in this manner (log of viscosity vs the inverse temperature) because this is the format of the viscosity plots of Bottinga and Weill (for example their Figure 7). We then used equation (1) to calculate the length of the flow in the same way as Figure 1 was calculated. We have used three different viscosities and two different gravities, and plotted the six lines in Figure 5b.

*Viscosity Error analysis*. In order to provide some measure of the error of these viscosity derivations, we use the error estimate from Anderson et al. (2022). The average Root Mean Squared Error of Prediction (RMSEP) from their Table 5 is 1.85 wt %. We also use the error estimate from Bottinga and Weill (1972) for their viscosity model. Figure 6 of Bottinga and Weill shows that 99% of their observations are within $-0.75 < \Delta\ln\eta < 0.75$ of their viscosity model. We therefore use +/- 0.75 as the error band for the ln viscosity.

Table 1. Elemental abundances and normative mineralogies obtained using SuperCam LIBS during the Crater Floor campaign (Wiens, 2022).

| Element oxide | Máaz | Artuby | Séítah |
|---|---|---|---|
| $SiO_2$ | 54.2 | 49.7 | 44.8 |
| $TiO_2$ | 0.6 | 0.9 | 0.2 |
| $Al_2O_3$ | 10.5 | 7.4 | 3.9 |
| FeO$_T$ | 20.2 | 25.3 | 22.6 |
| MgO | 2.7 | 3.8 | 21.4 |
| CaO | 5.5 | 7.7 | 3.6 |
| Na$_2$O | 3.3 | 2.7 | 1.3 |
| K$_2$O | 1.2 | 0.6 | 0.2 |
| Quartz | 5.6 | 0.0 | 0.0 |
| Plagioclase | 37.9 | 30.0 | 15.2 |
| Orthoclase | 7.1 | 3.6 | 1.2 |
| Diopside | 4.8 | 16.7 | 1.5 |
| Hypersthene | 37.9 | 41.5 | 34.7 |
| Olivine | 0.0 | 0.9 | 41.1 |
| Ilmenite | 1.1 | 1.7 | 0.4 |
| Magnetite | 1.6 | 2.0 | 1.8 |
| An # plag | 25.3 | 22.7 | 26.5 |
| Mg # | 19.3 | 21.1 | 62.8 |
| Density (g/cc) | 3.10 | 3.27 | 3.37 |
| | ±0.02 | ±0.03 | ±0.01 |
| Viscosity at 1470° K (6.8 inv. T) | 665.14 ($e^{6.5}$) | 492.75 ($e^{6.2}$) | 270.43 ($e^{5.6}$) |

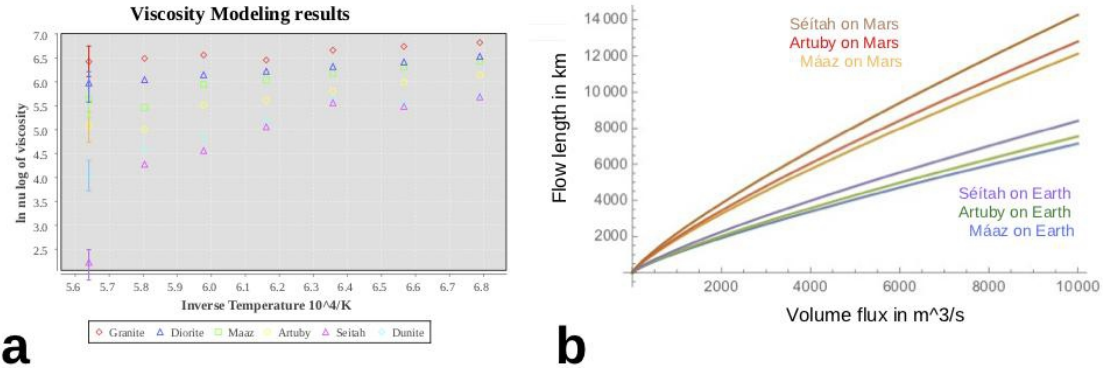

Figure 5. a.) Results of viscosity calculations for Máaz, Artuby, and Séítah rocks compared with terrestrial granite, diorite and dunite. Viscosity error bars are shown for the leftmost values and are the same for all points. b.) Flow lengths for viscosities of Séítah, Artuby and Máaz on Earth and Mars.

*3.2. Clay alteration*

*3.2.1 CRISM orbital maps*

To visualize the clays that are present in the olivine-clay-carbonate, we have produced a map of the orbital CRISM Half Resolution Long (HRL) image, HRL40FF, over Jezero crater (Figure 6). This includes the area of Mars 2020 operations, Octavia E. Butler Landing site, Jezero Delta, and the marginal carbonates region (Williford et al., 2018; Farley et al., 2020; Stack et al., 2020). Alongside a 0.905 μm channel image, we have included three absorption band maps from the CRISM image for 2.31 μm (clay and carbonate), 2.5 μm (carbonate) and 2.38 μm (clay). Here we focus on the clay signature; for more discussion of the carbonate spectral signature, see (Mandon et al., 2020; Brown et al., 2020; Zastrow and Glotch, 2021)**.**

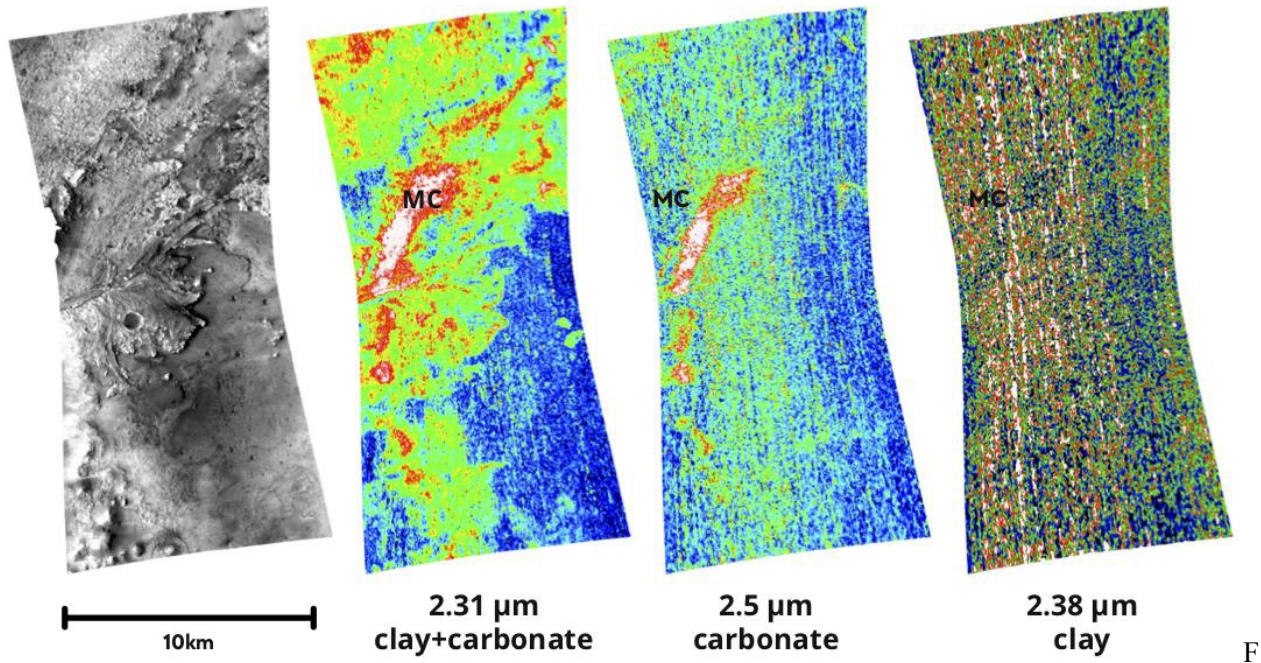

Figure 6. HRL40FF 0.905 μm map, 2.31 μm, 2.5 μm, 2.38 μm band amplitude maps. The location of the marginal carbonates is also shown as MC.

### *3.2.2 Correlation maps*

As can be seen from Figure 6, mapping the 2.38 μm clay band is very difficult due to it being close to the noise level. To overcome this and gain further insights into the presence of clays in the CRISM HRL40FF image, we have constructed a correlation map consisting of the 2.31, 2.38 and 2.5 μm bands. For each of these three band maps, we have the four parameters of the Asymmetric Gaussian band, meaning we have 12 parameters total. Figure 7 displays the correlation plots for HRL40FF for band depth amplitude (2.38)-amplitude(2.31)-amplitude(2.5). This can be interpreted as the *x* value being the amplitude of the 2.38 μm band, the *y* value as the amplitude of the 2.31 μm band, and the coloring of the points reflects the amplitude of the 2.5 μm band.

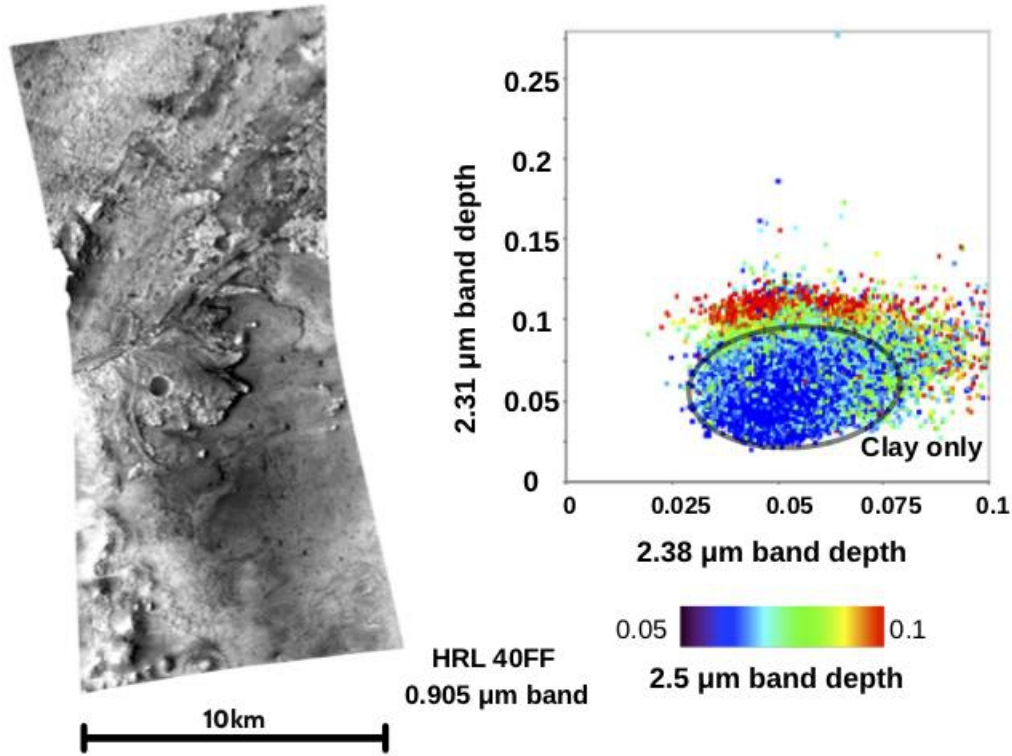

Figure 7. HRL40FF correlation plots of the 2.38 μm band depth (*x*-axis) vs. the 2.31 μm band depth (*y*-axis). Color indicates the 2.5 μm band depth, with red>green>blue.

These correlation plots show when all three bands are present and to what degree. Figure 7 shows that the points that are high in 2.5 μm amplitude are also high in 2.31 μm band amplitude. This allows us to be very specific and make a finding: there are regions of the HRL40FF image that display the presence of all three bands (known as the marginal carbonates, marked on Figure 6). However, there are regions where only clays are present, with only a weak to vanishing 2.5 μm carbonate band, indicated by blue colored points. This is marked "Clay only" on the Figure 7 correlation plot. This plot makes clear that the highest 2.5 μm band amplitudes are correlated with the highest 2.31 μm band amplitudes.

### 3.2.3 Venn diagrammatic approach

To get a quantitative estimate of the clay and carbonate present with the olivine in this image, we present a Venn diagram populated with data from HRL40FF where olivine, clay and carbonate are determined to be present. In order to construct this diagram, we used the following detection rules for the olivine, carbonate, and clay bands:

amplitude(olivine 1.0 μm band) > 0.29

amplitude(carbonate 2.5 µm band) > 0.08
amplitude(clay 2.38 µm band) > 0.065

For simplicity and reproducibility, these rules were run on the entire georeferenced HRL40FF CRISM image. No attempt was made (at first) to mask the image to regions of interest. We report the border pixels with no data as "edge pixels" and pixels with data as "ground pixels".

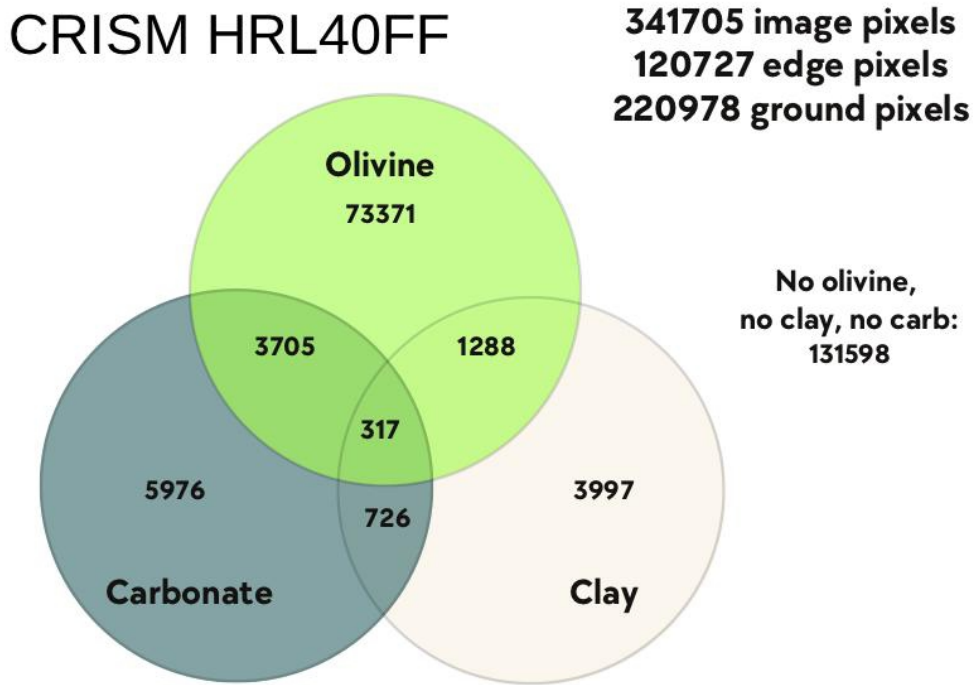

Figure 8. Venn diagram of olivine-clay-carbonate detections for HRL40FF showing overlapping pixels containing olivine, clay and carbonate. See text for details of calculation.

Figure 8 demonstrates that just more than half of the HRL40FF image is not olivine, clay or carbonate. Roughly one third of the ground pixels have olivine alone detected. "Pure" (i.e. without the other two minerals) clay and carbonate are present in roughly 3% and 2% of the pixels, with carbonate being more abundant than clay, using orbital data. This plot also demonstrates that there are large amounts of olivine without alteration, and small but significant amounts of clay and carbonate without significant olivine signatures in the HRL40FF scene.

In order to take a closer look at the Séítah region as seen from orbit, we have subset the HRL40FF image to a smaller image of only 1750 pixels covering the Séítah region. Figure 9 displays the Venn diagram analysis for just this Séítah subset. This shows that the subset CRISM image mostly contains olivine (~95%) and 2% of the pixels containing clay and 1.5% are carbonates and olivine. Relative to the whole 40FF image, the Séítah subset contains far more pure olivine, but relatively smaller amounts of clay and carbonate. This reflects the fact that the Séítah region carbonate and clays are relatively difficult to detect from orbit. In the section that follows, we supplement these orbital data with in situ data from the Mars 2020 rover.

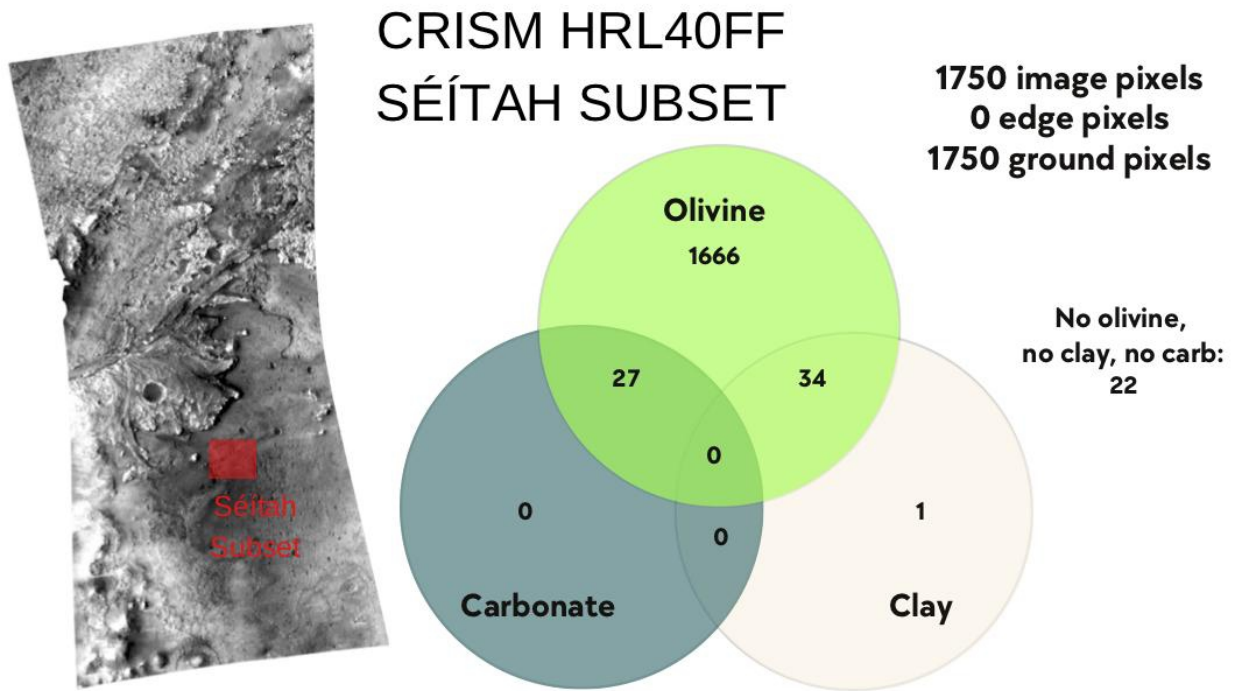

Figure 9. Venn diagram of olivine-clay-carbonate detections for the Séítah subset of HRL40FF, which is shown in a red box in the image at left. showing overlapping pixels containing olivine, clay and carbonate. See text for details of calculation.

### 3.3.1 VISIR In situ results

Figure 10 shows the VISIR spectra from the abraded patch Garde taken on Sol 207 from the Séítah formation. Absorption bands at 2.28, 2.33, 2.38 and possibly 2.46 µm were detected. The lack of a moderate strength 2.5 µm band suggests that this sample is not carbonate bearing; however, see the analysis by Clave et al. (2022) in which LIBS and Raman techniques suggest small amounts of carbonate are present, and Corpolongo et al. (2022) in which SHERLOC Raman analyses of the abrasion patch indicate carbonate. The presence of bands at 1.9, 2.28, 2.31, 2.38 and possibly at 2.46 µm in the VISIR indicates that the Garde sample is clay bearing. In addition, although weak, there is a band present in the spectra at 2.1 µm, and though this is a little offset (2.09 µm), it is also present in talc. Finally, the 2.46 μm band in serpentine is far weaker than in talc. Figure 10 also shows the example of a spectrum from Dourbes_Tailings_255 which shows the most convincing example we have found of bands at 1.94, 2.28, 2.31, 2.38 and at 2.46 μm, thereby constituting the best evidence in the Séítah formation for the presence of talc.

Also shown in Figure 10 are talc and serpentine (lizardite) spectra from Figure 4 for direct comparison with the band positions of the VISIR data.

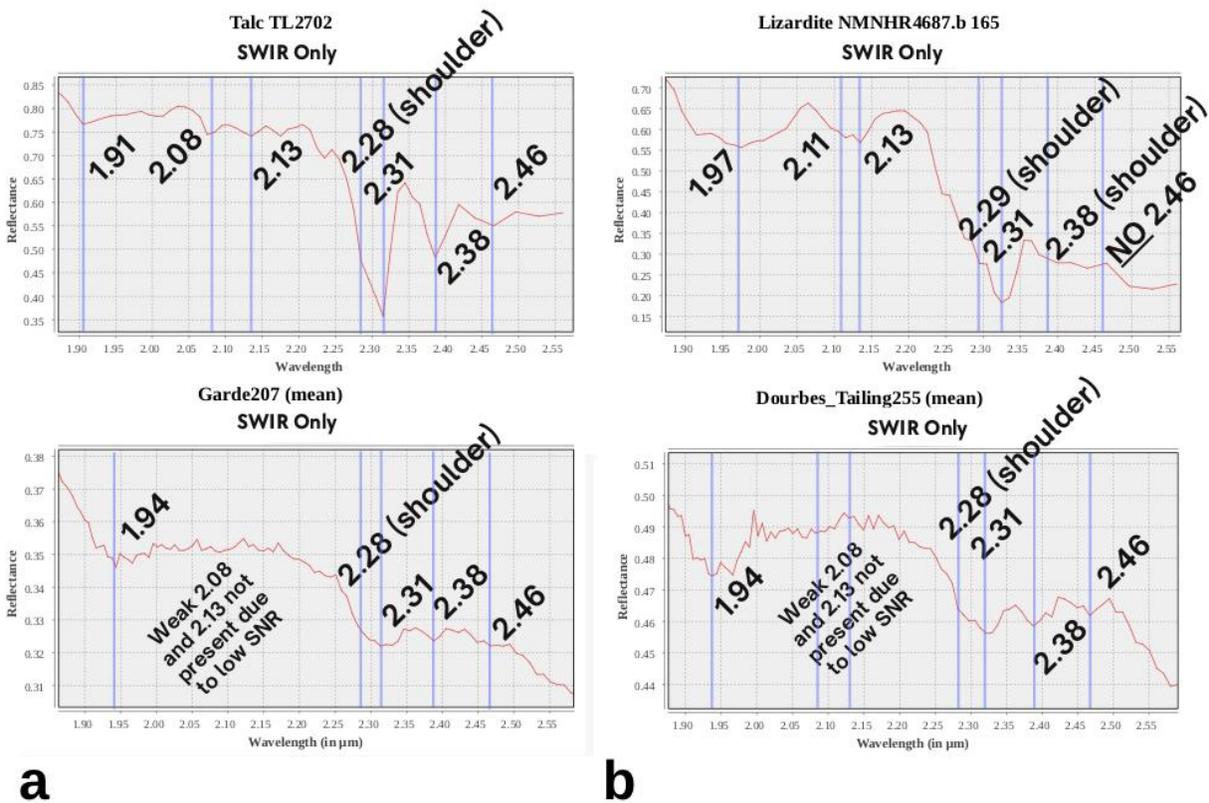

Figure 10. VISIR spectra of a. Garde_207 and b. Dourbes tailings from Sol 255 and plot of the USGS spectroscopy library spectrum of talc and lizardite from Figure 4.

*3.3.2 VISIR, RMI and Mastcam-Z tailings observations*

Figure 11 shows the MastCam-Z image for the Dourbes target (abraded rock), showing the tailings that were captured by the VISIR instrument. It should be noted that the tailings are intensely bright, especially compared with the surrounding bedrock, which is again consistent with the presence of talc, which has a high reflectance in the visible region. It should be noted that a decrease of grain size will also increase the albedo, so we do not consider the brightness change to be definitive evidence for talc over serpentine.

In Figure 4 we presented USGS library spectra of talc and lizardite. For these samples, it can be noted that the albedo of talc is uniform in the visible range, which is responsible for its white appearance. It is intensely bright at 0.9 µm across the visible range. The albedo of the lizardite peaks at 0.7 µm and has a positive slope across the visible which reflects its green appearance. Again, the bright nature of the tailings pile is suggestive of the presence of talc rather than lizardite.

Also shown in Figure 11 are the spectra from the VISIR observations of targets Dourbes, Salette and Brac, which are shown with the spectra of the original rock surface and the tailings after abrasion. The key thing to note, particularly in the Dourbes example, is the relative overall increase in albedo, and the appearance of a band at 2.46 µm which is not present (or only weak) in the original rock. This is strongly suggestive that the tailings have made the presence of the clay more obvious in the spectra, and because the 2.46 µm band is not present in lizardite (Figure 10), this strongly suggests the tailings contain the mineral talc.

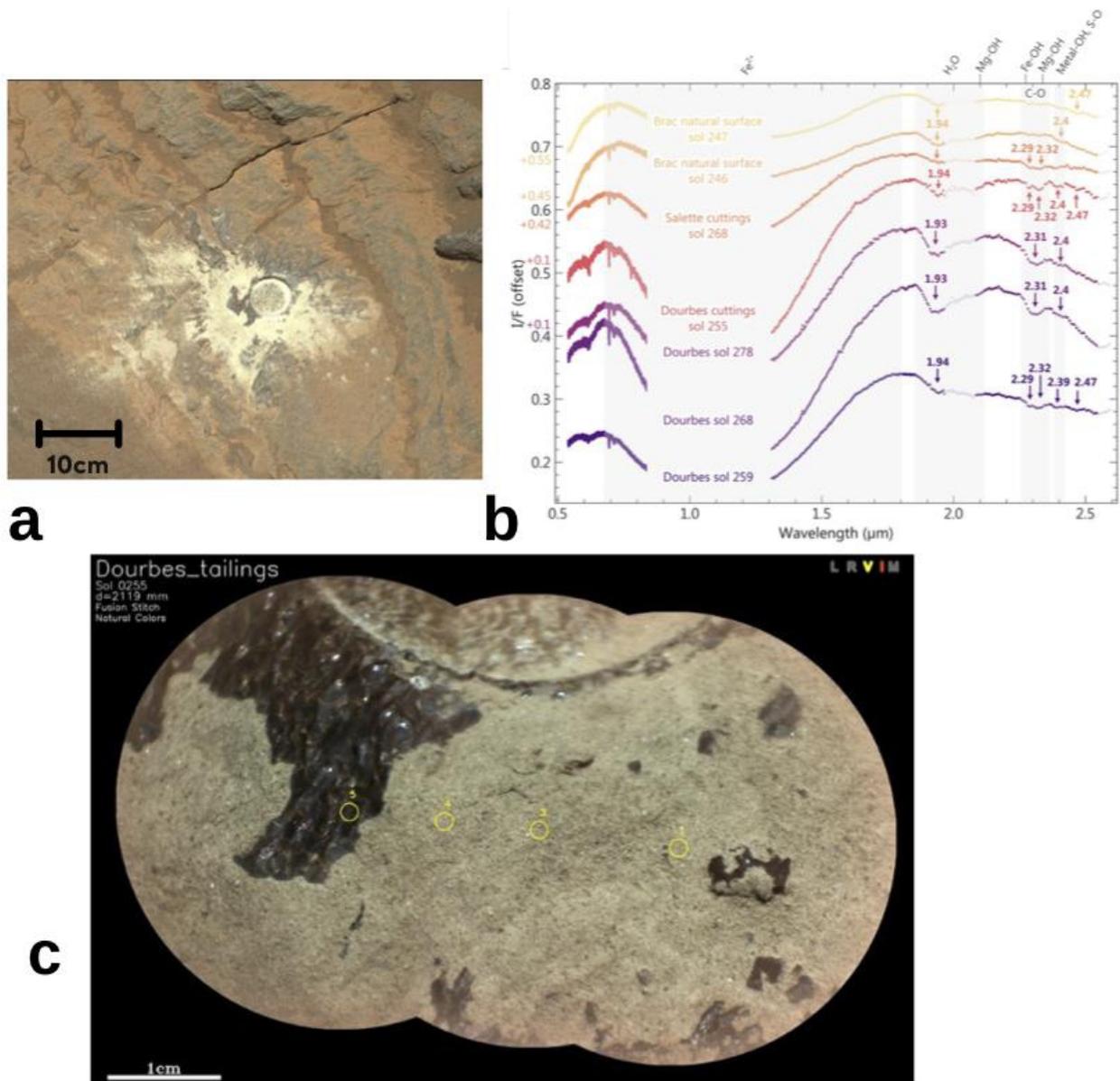

Figure 11. a.) Mastcam-Z of Target Dourbes with tailings on Sol 255 after abrasion in the Brac workspace. Note the intense brightness of tailings compared to the original rock surface. For scale, the abrasion patch is 5cm across. Mastcam Z image id number ZR0_0255_0689576242_738EBY_N0080000ZCAM08278_0340LMJ
b.) Spectra of the tailings compared to the original rock surface. Note the presence of the 2.46 μm band in the tailings spectra c.) SuperCam RMI image of target Dourbes tailings after abrasion. The numbered yellow circles indicate the locations at which the data was taken and their fields of view (65% of total signal encircled).

### 3.3.3 SuperCam LIBS ternary plot

Figure 12 shows a Si+Al-Fe+Mg-Ca+Na+K ternary diagram displaying the SuperCam LIBS elemental composition results for the 10 points measured on the target Ubraye_255 which is a rock showing cumulate texture located near the Dourbes abrasion patch in the Brac workspace (Figure 1c). The elemental composition lies very close to the olivine calibration target and also overlaps with the region of talc and serpentine. Given how close these points are and the errors inherent in the planetary LIBS observations (Anderson et al., 2022) it can become challenging to tell these mineral compositions apart by just looking at the major elements. We provide in the supplementary section S1 a discussion of errors and discrimination of points in the ternary plot.

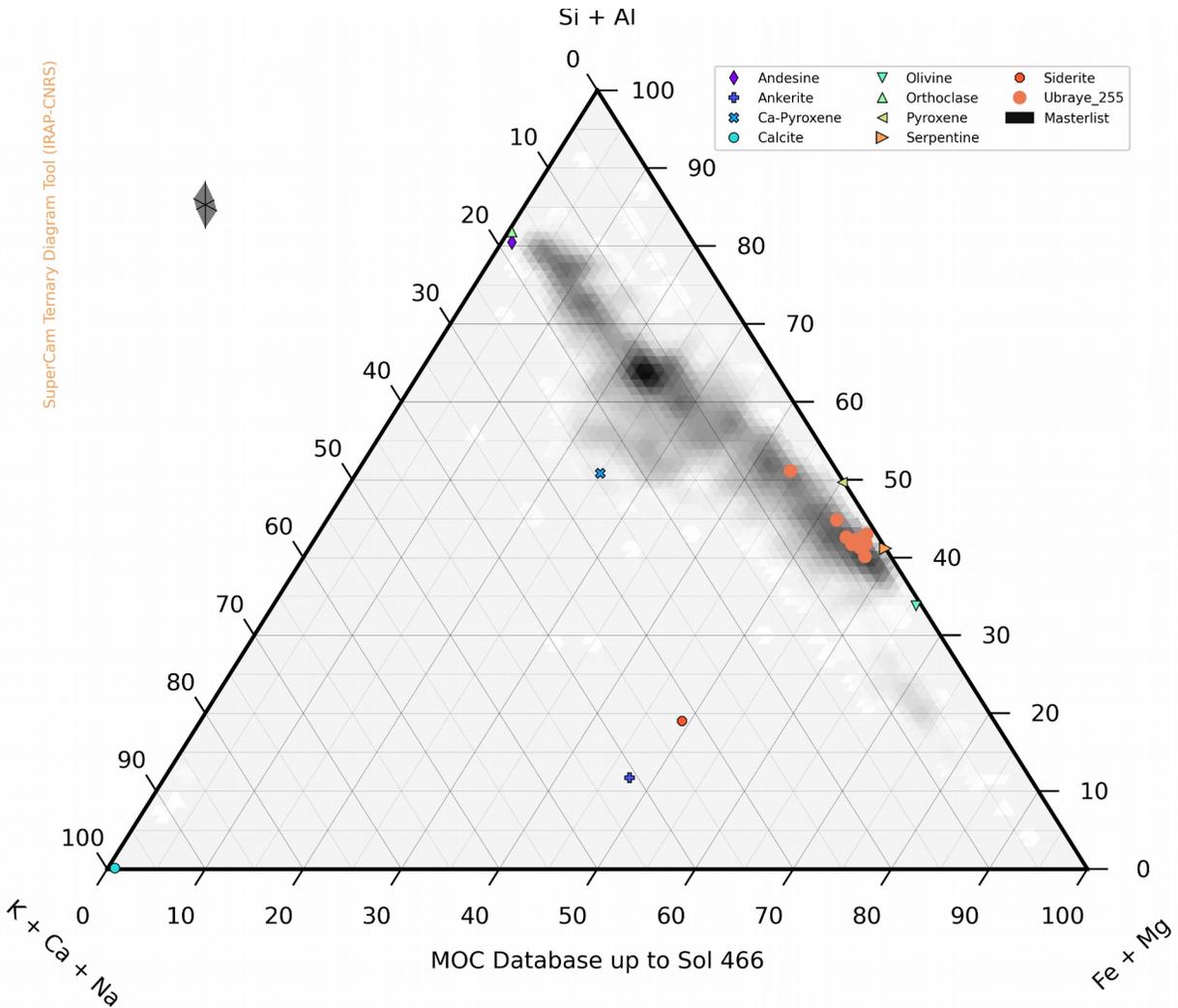

Figure 12. LIBS and geochemistry ternary molar plot indicating Ubraye_255 (orange circles without outline) cumulate target showing overlapping olivine (green inverted triangle) and serpentine (right-pointing orange triangle) standards on board the rover. The gray polygon at top left above the main figure indicates the error bars of the plot.

Table 2 gives an example composition from Cine and Ubraye showing the very low (<4 wt %) Al present in these samples. Anderson et al. (2022) Table 5 gives the mean accuracy of the $Al_2O_3$ estimate at +/- 1.8wt %. For these LIBS targets (Cine/Garde, Udraye/Dourbes which are all in the Séítah region) we can eliminate the possibility of all Al-bearing clays such as saponite, smectite and montmorillonite.

Table 2. SuperCam LIBS Elemental abundances from targets Cine_206 (cumulate rock next to Garde in Bastide workspace) and Ubraye_255 (cumulate rock near Dourbes in Brac workspace). Cine_206 point 3 and 5 and Ubraye_255 points 1-4 and 6-10 show a very low amount of Al to within error (<4%) in the cumulate textured rocks. Green boxes highlight the high values of $FeO_T$ and MgO and red boxes highlight the low values of $Al_2O_3$. S.D. is the standard deviation for that elemental data.

| Target | # | $SiO_2$ | S.D. | $TiO_2$ | $SiO_2$ Std Dev | $Al_2O_3$ | S.D. | $FeO_T$ | S.D. | MgO | S.D. | CaO | S.D. | $Na_2O$ | S.D. | $K_2O$ | S.D. | Total |
|---|---|---|---|---|---|---|---|---|---|---|---|---|---|---|---|---|---|---|
| Cine | 1 | 44.99 | 1.19 | 0.43 | 0.02 | 3.27 | 0.53 | 13.37 | 1.52 | 11.54 | 1.16 | 15.84 | 0.68 | 0.39 | 0.13 | 0 | 0.17 | 89.83 |
| Cine | 2 | 51.05 | 1.9 | 0.04 | 0 | 2.94 | 0.97 | 25.74 | 2.06 | 13.4 | 1.96 | 4 | 0.5 | 1.19 | 0.38 | 0 | 0.15 | 98.36 |
| Cine | 3 | 45.13 | 2.03 | 0.01 | 0.02 | 0.87 | 0.95 | 34.59 | 2.13 | 28.93 | 1.27 | 1.81 | 1.09 | 0.37 | 0.33 | 0 | 0.05 | 111.71 |
| Cine | 4 | 43.39 | 1.15 | 0.45 | 0.03 | 3.91 | 1.34 | 10.43 | 2.41 | 11.13 | 0.92 | 13.72 | 0.79 | 0.49 | 0.17 | 0 | 0.19 | 83.52 |
| Cine | 5 | 45.34 | 2.14 | 0.01 | 0.01 | 2.89 | 1.08 | 36.37 | 3.56 | 28.35 | 2.22 | 1.28 | 0.48 | 0.51 | 0.11 | 0 | 0.03 | 114.75 |

| Target | # | $SiO_2$ | S.D. | $TiO_2$ | $SiO_2$ Std Dev | $Al_2O_3$ | S.D. | $FeO_T$ | S.D. | MgO | S.D. | CaO | S.D. | $Na_2O$ | S.D. | $K_2O$ | S.D. | Total |
|---|---|---|---|---|---|---|---|---|---|---|---|---|---|---|---|---|---|---|
| Ubraye_255 | 1 | 48.72 | 2.07 | 0.01 | 0 | 2.02 | 0.88 | 24.5 | 2.92 | 31.5 | 0.85 | 1.43 | 0.54 | 1.22 | 0.21 | 0 | 0.23 | 109.4 |
| Ubraye_255 | 2 | 44.76 | 2.2 | 0.01 | 0 | 2.25 | 0.88 | 25.94 | 2.24 | 27.91 | 2.7 | 1.25 | 0.37 | 0.34 | 0.09 | 0 | 0 | 102.46 |
| Ubraye_255 | 3 | 51.41 | 2.57 | 0.02 | 0.01 | 1.57 | 1.06 | 23.63 | 3.05 | 32.07 | 0.85 | 0.8 | 0.4 | 1.37 | 0.17 | 0.49 | 0.41 | 111.36 |
| Ubraye_255 | 4 | 45.97 | 1.52 | 0.01 | 0 | 1.79 | 0.75 | 26.35 | 2.87 | 28.87 | 2.03 | 1.32 | 0.36 | 0.3 | 0.1 | 0 | 0.03 | 104.61 |
| Ubraye_255 | 5 | 49.34 | 2.74 | 0.04 | 0 | 3.92 | 1.14 | 28.53 | 3.78 | 15.25 | 2.62 | 1.65 | 0.44 | 1.32 | 0.14 | 0.52 | 0.38 | 100.57 |
| Ubraye_255 | 6 | 46.36 | 1.4 | 0.01 | 0 | 1.78 | 1.11 | 21.35 | 1.86 | 30.32 | 1.16 | 0.56 | 0.21 | 0.29 | 0.05 | 0 | 0.15 | 100.67 |
| Ubraye_255 | 7 | 43.03 | 7.28 | 0.03 | 0.01 | 2.59 | 1.32 | 27.74 | 3.44 | 26.59 | 0.89 | 1.52 | 0.42 | 0.65 | 0.13 | 0 | 0.12 | 102.15 |
| Ubraye_255 | 8 | 47.04 | 1.74 | 0.01 | 0 | 1.67 | 0.72 | 23.18 | 2.38 | 29.89 | 1.04 | 1.41 | 0.34 | 0.65 | 0.17 | 0 | 0.07 | 103.85 |
| Ubraye_255 | 9 | 45.92 | 2.24 | 0.49 | 0.11 | 3.47 | 0.79 | 25.17 | 2.79 | 24.79 | 0.66 | 1.53 | 0.3 | 0.97 | 0.18 | 0 | 0.04 | 102.34 |
| Ubraye_255 | 10 | 45.84 | 2.18 | 0.01 | 0.01 | 0.54 | 0.67 | 27.48 | 2.68 | 28.97 | 2.2 | 1.46 | 0.21 | 0.76 | 0.15 | 0 | 0.03 | 105.06 |

## 4. Discussion

With the insights now gained from bringing together orbital and in situ observations, we now discuss two aspects of the geological history of the olivine-clay-carbonate lithology.

*4.1 Emplacement of olivine-clay-carbonate lithology and exposure today*

Flood lavas (e.g. the Columbia River basalt) are characterized by 1) large spatial extent, 2) low viscosity, 3) relatively thick (5-45 m) flows 4) fed by fissure eruptions which are covered by the erupting lava (Plescia, 1990). This is in contrast to (or could evolve over time to) plains lavas (e.g. Indian River basalt), which have smaller-extent, higher-viscosity, thinner flows accompanied by low shields.

Given the low viscosity for the lithology calculated in this paper, the results of RIMFAX showing the relatively thick flows, and the lack of low shields in the region from orbital mapping, we consider it likely the olivine-clay-carbonate at the Séítah formation was emplaced as a flood lava. Future observations at the Marginal Carbonates and outside Jezero will test what other formation mechanisms are involved in other parts of the lithology.

Under the flood lava hypothesis, the olivine-clay-carbonate lithology was likely variable in thickness, and in catchment regions of deep relief, such as Jezero crater, the flow formed a lava lake several decimeters thick.

*Effect of viscosity.* Worster et al. showed that by building a model of heat flow in a lake cool from above and insulated (or cooled) from below, and utilizing the balancing of the heat equation with crystallization of the melt, they could estimate the total crystallization amount. Their results are shown with their model in Figure 3. Figure 3b also shows that for inferred viscosities typical of the Séítah elemental composition, using the two base cooling scenarios as error bounds, the formation of crystals could have filled up to ~40%+/-10% of the volume of the lake.

This lava lake has then been eroded back to reveal the cumulate portion of the lava lake and then covered by crater fill consisting of igneous material from later eruptions such as the Máaz formation (Udry, 2022) and Artuby member of the Máaz unit (Alwmark, 2022) which has itself then been eroded back in locations to reveal the olivine-clay-carbonate beneath it (Hiesinger and Head, 2004; Goudge et al., 2015). Calef et al. (2022) report pahoehoe lava ejecta blocks associated with Adziilii crater. Ground-penetrating radar imaging performed by RIMFAX in the vicinity of Adziilii crater reveal sub-horizontal high-amplitude reflectors at ~ 6 m burial depth including ravioli-shaped reflector geometries interpreted to represent cross-sections of pahoehoe flow lobes (Calef, 2022). RIMFAX imaging around Adziilii crater indicates the thickness of the Máaz layer to be ~2 m. The base of the Séítah unit currently remains unclear, but is at least ~15 m below surface at the deepest locations observed so far by RIMFAX.

Assuming for the sake of the argument that the Séítah formation was 10 m deep if it was formed as a lava lake, this suggests (using Figure 3b) that approximately the lower ~4+/-1m consists of the cumulate Bastide workspace (Figure 1b) in which Cine_206 and Garde_207 were observed.

*4.1.1 Spatial distribution of lithology as flood lava*

Wilson and Head (1994) pointed out that lava flows on Mars will extend roughly six times further than lava of the same composition on Earth. This is due to the lower gravity and the lower pressure of the current Martian atmosphere. A higher pressure Noachian atmosphere would decrease the runout. The emplacement of the lava that created the Séítah formation olivine cumulate covered a vast region of the planet (Figure 13). Kremer et al. (2019) mapped the unit to extend over both sides of the Nili Fossae, and showed the unit to be predominantly exposed on the eastern side of the Nili Fossae, and truncated by Syrtis Major to the south, and later cratering activity, most notable by Hargraves Crater in the middle of the two predominant Fossae. Considering only the region of the Isidis Planitia basin complex currently covered by the olivine-clay-carbonate unit, and excluding the olivine on the Libya Montes region on the southern edge of the basin, this area extends roughly 500 km north to south and 350 km east to west and is approximately the size of Florida in the United States (~170,000 km$^2$).

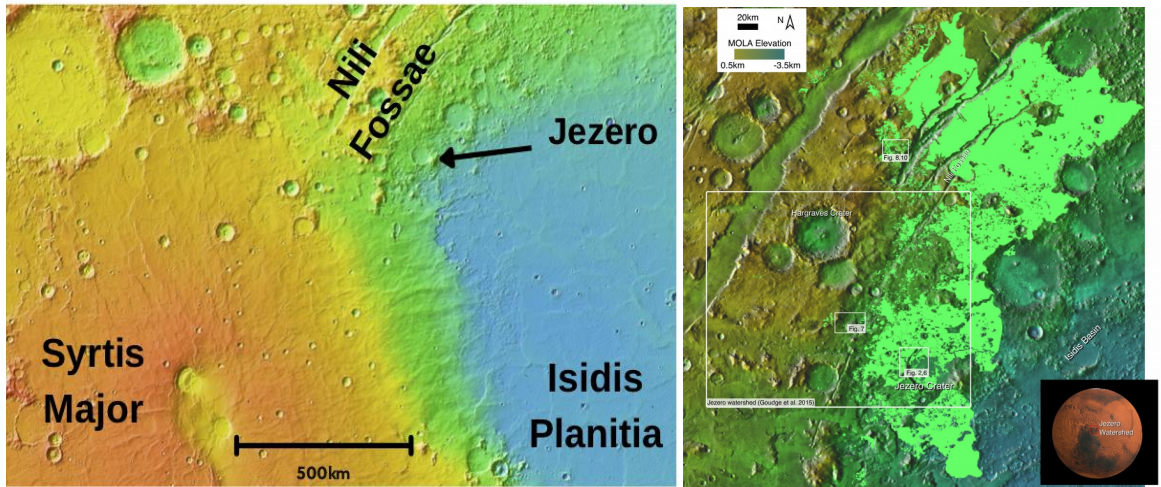

Figure 13. a.) MOLA image showing the relationship of western edge of the Isidis Basin complex, giving the relative positions of Syrtis Major, Nili Fossae and Jezero Crater. b.) Figure 1 of Brown et al. (2020) showing the extent of the olivine-clay-carbonate lithology as mapped by Kremer et al. (2019) in a region about the size of Florida. Note obscuration by later crater ejecta blankets, especially Hargraves Crater.

The direct contact between the Syrtis Major lavas and the olivine-clay-carbonate is striking and leads one to conjecture whether the Fossae extends beneath the Syrtis Major construct; although this has not been previously addressed, we will now postulate about how this geophysical relation might have occurred. Rampey and Harvey 2012 identified the wide depressed footprint of Syrtis Major and suggested that:
> "early eruptions at Syrtis Major might have consisted of voluminous low-viscosity, possibly komatiitic magmas, whose powers of thermal erosion … may have been responsible for the wide, depressed footprint of Syrtis Major."

And at the end of the same paragraph:
> "... the youngest and most visible surface rocks of which would be basaltic komatiites, but whose earlier, more voluminous rocks would have been ultramafic. In the absence of tectonic effects these deep-seated rocks would have restricted exposure."

We would contend that the missing earlier, more voluminous ultramafic rocks, are, in fact, the olivine-clay-carbonate lithology we have been studying.

Somewhat more speculatively, we might make some passing comments which may be useful for the collective consciousness to test in future investigations. The suggestions are four-fold and currently have no basis in observation, but are consistent with observations so far and may be tested in future observations, for example when the *Perseverance* rover can inspect the Nili Fossae in situ.

1) The fissures that were the source of the olivine-clay-carbonate lava were possibly linked to, or in fact used, the Nili Fossae fractures on the northwest of Isidis Planitia, and opened after the Isidis impact (Wichman and Schultz, 1989). The analogy might be made to the relationship between Cerberus Rupes and Cerberus flood lavas (Plescia, 1990).

2) At Cerberus, it is possible that the flood lava evolved to a plains style lava. This may also have happened in the Isidis basin, where Syrtis Major may have been created after fissure lavas became more viscous, and the Syrtis shield may now cover fissures from an earlier phase of flood volcanism that produced the olivine-clay-carbonate lithology. Figure 13 shows a Mars Orbiter Laser Altimeter (MOLA) altimetry map of the region highlighting the relationships between the Isidis, Syrtis, Nili Fossae and Jezero crater. Rampey and Harvey (2012) suggested that Syrtis Major evolved from ultramafic to mafic over time and that super-hot komatiite lava may have eroded basement rocks. As noted by previous authors, no olivine has been found to date in the Fossae floor or walls (Hamilton and Christensen, 2005; Mustard et al., 2007). However, it is possible that evidence of this olivine may yet be found buried at depth by Syrtis Major lavas. Channelized lava flows travel much further than non-channelized flows (Pinkerton and Wilson, 1994), and the last Syrtis Major lavas may have traveled more easily in the Nili Fossae channels.

3) At Cerberus, the lavas are much younger than the olivine-clay-carbonate, post-date water-related fluid channels, and are often guided by those channels. Volcanic activity commenced at Syrtis between 3.6-3.9 Ga (Robbins et al., 2011) around the same time as the oldest deposits on Tharsis were emplaced at 3.67 Ga (Isherwood et al., 2013). The age of the olivine-clay-carbonate (3.82 +/- 0.07Ga, Mandon et al. (2020)) would imply that no surface water channels were available to confine the flow, but might instead be affected by the widespread emplacement of the Noachian basement rocks.

4) During emplacement of this super-hot lava lake, it might be possible that Séítah sank into and eroded the regolith it was laid down upon (Huppert et al., 1984). Evidence for all four of these suggestions might be found in exposures sounded by RIMFAX during the Mars 2020 mission in the Outside Jezero part of the mission.

*4.1.2 Properties of cumulate layer and terrestrial analog*

Huppert and Sparks (1981) discuss the manner in which an olivine bearing ultramafic cumulate layer might form in a magma body beneath a less dense plagioclase bearing basaltic layer. This model is a reasonable candidate for the formation of the Séítah formation and the Artuby member of the Máaz formation (see also Crumpler et al., (2022)). It may even be possible that the lava which formed Séítah was contaminated during eruption and emplacement (Huppert and Sparks, 1985), and given its low viscosity and high temperature, this may have differentiated the upper Issole member as a contaminated version of the lava producing the lower cumulate Bastide member (Crumpler, 2022).

Figure 14 compares the Martian olivine cumulate target Cine_206 and the terrestrial olivine cumulate target AJB0503100. Both show equant (1 to 1.5 mm diameter) grains that are interlocking. Both were likely formed as the basal part of a lava flow or ponded lake when the olivine crystals fell out of suspension and collected at the basal cumulate layer. The terrestrial example on the right was sampled from the talc-carbonate altered unit at the bottom of the Mt Ada Basalt in the Dresser Formation of the North Pole Dome in the Pilbara region of Western Australia. The sample is part of a komatiite sequence and specifically the B2 adcumulate layer (Arndt et al., 2004). We have shown in this paper that the target Cine shares many spectral, mineralogical and morphological characteristics of the terrestrial target. Although a Martian lava lake is likely to have formed from a cooler mantle source than the terrestrial komatiite (Putirka, 2016), we contend that it is the viscosity we have derived here that governs emplacement properties of the lava flow. Direct comparison of the viscosity of the two units is challenging due to the high Mg nature and alteration of the terrestrial sample (MgO 18.9 wt.%, $FeO_T$ 7.6 wt.% from microprobe, presented in Brown (2006)) versus the higher Fe content of Target Cine ($SiO_2$ 45.24 wt %, MgO 28.64 wt.% and $FeO_T$ 35.48 wt.%, which is the average of Cine points 3 and 5 from Table 2). It should be noted that Cine does not have a true bulk rock content because it is a cumulate rock of material that was sourced from elsewhere, most likely within the upper zone of the lava lake.

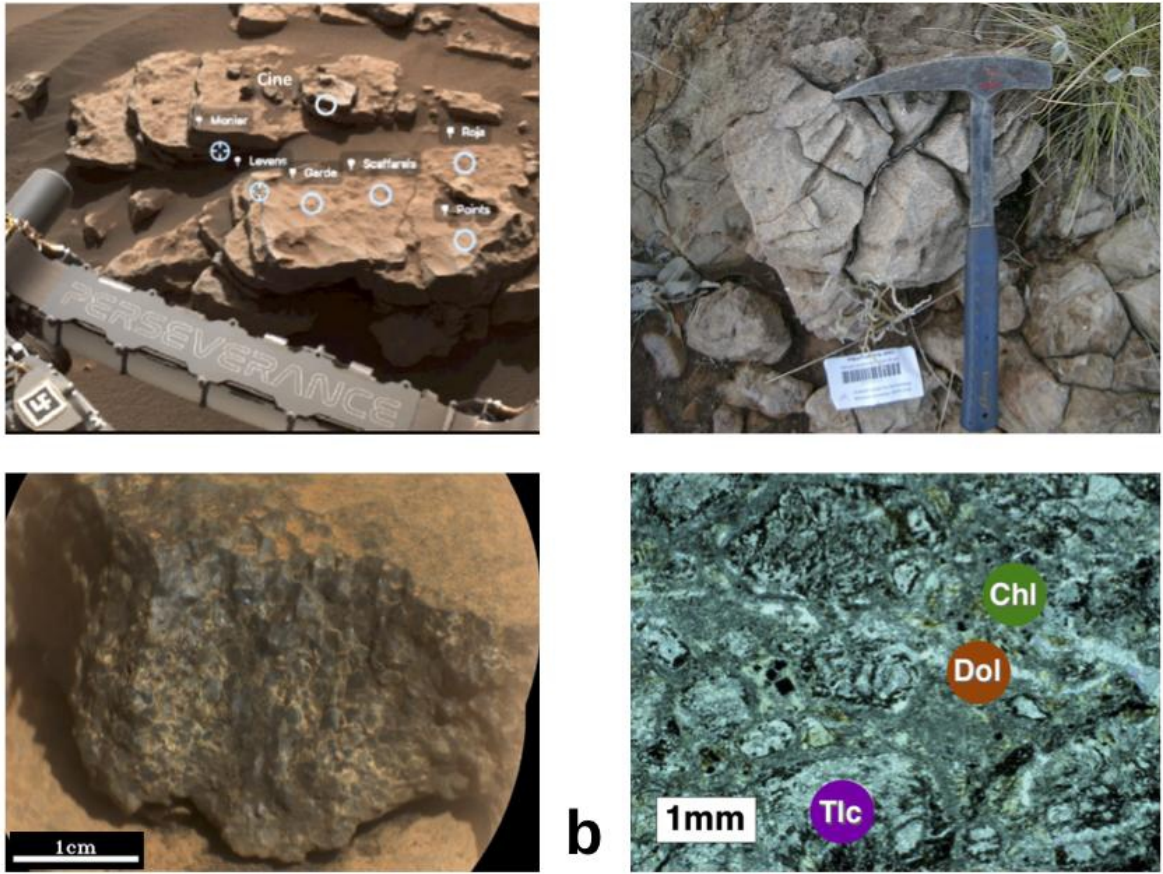

Figure 14. a.) (bottom) SuperCam RMI of target Cine showing olivine cumulate texture at mm scale. (top) Bastide workspace, location of Cine relative to Garde, rover arm for scale. b.) (bottom) Thin section of talc-carbonate sample AJB0503100 from Brown (2005; 2006) with mm size talc (Tlc) replacing olivine, dolomite (Dol) and chlorite (Chl) identified using electron microprobe. (top) Context image for target AJB0503100 with hammer for scale.

*4.2 Post emplacement clay-carbonate alteration*

The fact that ~3.94 Ga olivine is still well preserved on Mars and relatively unaltered speaks volumes to the amount of alteration and volatiles present on the fourth planet. In terrestrial greenstone Archean terranes, olivine is usually only present as a polymorph of its former self, often replaced by clays such as serpentine or talc (Figure 14b).

Wilson and Head (1994) suggest that lavas on Mars have fewer volatiles than lavas on Earth. Despite this, the low amount of alteration present means it is still possible that the alteration fluids were juvenile water and $CO_2$, sourced from the fissure, as suggested and discussed by Brown et al. (2020).

The limited extent of post emplacement clay and carbonate alteration is evident from the Venn diagrams we have presented in Figure 8 and 9, and is also the subject of two studies, including clay (Mandon, 2022) and carbonate (Clave, 2022). In a recently published paper by Mandon et al. (2022), it was shown that in olivine-carbonate-serpentine mixtures, the bands of serpentine are deep for percentages < 5 wt.% of serpentine. This also has implications for the alteration history, as the rocks in the Séítah formation might not have extensively interacted with fluid compared to some other parts of the regional olivine unit, which appear more altered from orbit.

*4.3 Other emplacement models*

There are a number of aspects of the regional morphology and properties of the olivine-clay-carbonate lithology that remain poorly understood, including draping morphology and laterally continuous layering and lack of clear lobate features. These have been used to argue in favor of an explosive volcanic origin for the lithology (Rogers et al., 2018; Kremer et al., 2019; Mandon et al., 2020; Ruff et al., 2022). Ravanis et al. (2022) report on modeling to determine grain sizes for Martian Nili Fossae ashfall deposits in order to better understand the potential for pyroclastic fall deposits to be emplaced away from the eruption source. Wilson and Head (1994) modeled clasts with diameters between 100 µm and 10 mm and maximum travel ranges of pyroclasts in a current day (6 mbar) Martian atmosphere for eruption cloud heights from 50-300 km. Their Figure 21 demonstrated that the maximum range drops sharply as grain size increases. They also presented a modeled maximum pyroclastic flow run out length as a function of eruption velocity for three different surface friction coefficients. Their Figure 23 shows that the 1.5 mm olivine grain size would only travel about 20 km away from the vent through pyroclastic fall, and thus far, the Mars 2020 team is yet to discover pyroclastic signatures in the pictures of abrasion patches obtained by the rover.

*4.4 Future prospects and Mars Sample Return*

At the time of writing this paper and other accompanying papers in the Jezero Crater Floor collection, the ongoing progress of the *Perseverance* rover has brought it to the Jezero Delta campaign of its mission. Within the delta, the delta curvilinear unit (*DCu*) and some parts of the delta have been documented to contain carbonates, from orbital studies (Goudge et al., 2015, 2018), and are now being characterized by surface observations (Mangold et al., 2021). We anticipate unexpected findings in the rocks and cobbles of the delta, which will provide us views of the outside watershed at irregular intervals. Figure 15 gives the timeline of events in Martian geological history that are critical to understanding its evolution and reliance upon regional and global events. We point out that this timeline shows an approximate correlation between the Jezero Delta and the recently hypothesized second generation ocean, called Deuteronilus, aged approximately at 3.65Ga, and discussed in Citron et al. (2018) which they hypothesized filled Isidis with water. The existence and timing of these oceans is a source of ongoing debate (Sholes et al., 2021; Sholes and Rivera-Hernández, 2022) and during the Delta campaign we may hope to deepen our understanding of the source of the water that once filled Jezero.

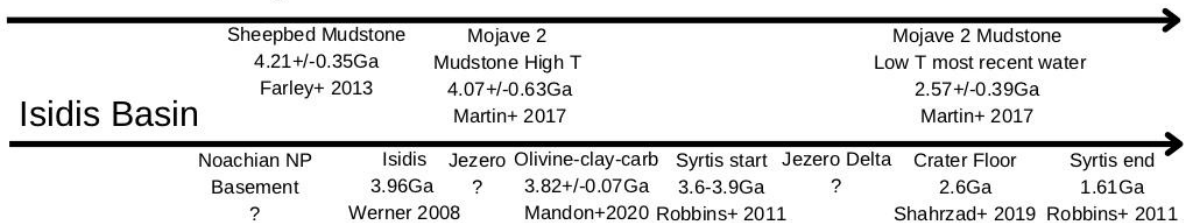

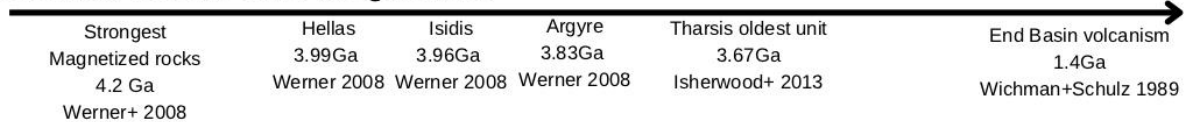

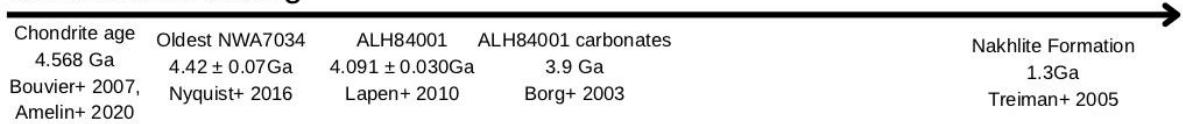

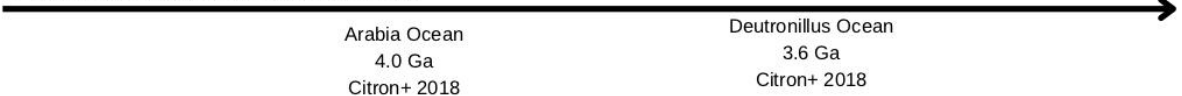

Figure 15. Timeline of events relevant to the emplacement and retention of the olivine-clay-carbonate lithology. Dates are from various sources (Amelin et al., 2002; Borg et al., 1999; Citron et al., 2018; Farley et al., 2013; Isherwood et al., 2013; Lapen et al., 2010; Mandon et al., 2020; Martin et al., 2017; Nyquist et al., 2016; Robbins et al., 2011; Shahrzad et al., 2019; Treiman, 2005; Werner, 2008; Wichman and Schultz, 1989).

After the Delta campaign, the rover will pass either nearby or through the marginal carbonates, which are the greatest concentration of carbonates at Jezero (Figure 6).

Throughout its mission *Perseverance* is collecting samples for later return to Earth as part of the Mars Sample Return Mission. The samples collected in the Crater Floor campaign are discussed in Simon et al. (2022). Upon return to Earth, we anticipate these activities to test the postulates and expand the findings of this paper:

- Use of radiometric age dating on the grains and mesostasis,
- Determination of clay and carbonate type using Microprobe and X-ray Diffraction,
- Determination of clay and carbonate relative timing using X-ray tomography,
- Source of carbonate using stable isotopes,
- Age dating of delta topsets and bottom sets samples to constrain the age of the Jezero Delta.

## 5. Conclusions

The *Perseverance* mission completed the Crater Floor science campaign in April 2022, and thus the observations and interpretations we have reported here are limited to the olivine-clay-carbonate as it appears at the Séítah formation. We can infer characteristics of the unit beyond Jezero based on orbital observations (Goudge et al., 2015; Horgan et al., 2020; Mandon et al., 2020; Brown et al., 2020). For example, one observation of the orbital data demonstrated that the largest olivine spectral features (and inferred grain size) are found outside, and to the north, of Jezero (Brown et al., 2020). However, our findings for this regional lithology will remain uncertain until we are able to reach these locations for in situ observations. These occasions will no doubt be as informative as the visit of *Perseverance* to Séítah.

We have made the following new findings and interpretations:

1. We have used a Venn-diagrammatic approach to establish that the previously identified "olivine-carbonate" lithology should actually be called the "olivine-clay-carbonate" lithology.
2. Again using the Venn-diagrammatic approach, we have demonstrated the limited extent of alteration by clay and carbonate, particularly for the Séítah subset of HRL40FF.
3. We propose that the olivine-clay-carbonate lithology at the Séítah formation was emplaced as a flood lava, with low viscosity, relatively deep flows, and lack of low shield volcanoes.
4. We have presented a model suggesting that a thermal-convection-driven

cooling, balanced by crystallization of the lava, must have taken place in order to form large mm-sized crystals seen in the Séítah formation.
5. We have presented a model where flood lava ponded into a lava lake in Jezero crater, and then 40+/-10% of the magma in the lake crystallized. Based on RIMFAX data suggesting the Séítah formation was approximately 10m thick, this would lead to the bottom ~4m of the Séítah unit consisting of the Cine and Garde Bastide workspace olivine cumulate outcrop sampled by *Perseverance*. The original magma body would necessarily have been thicker to allow differentiation of the olivine crystals and settling to the bottom of the chamber. This upper part would be more pyroxene and olivine-rich, as described in Wiens et al. 2022. It is likely that this portion of the magma lake was removed prior to emplacement of later lava flows.
6. As seen in Figure 6, there are regions of the CRISM HRL40FF image that display the presence of all three olivine, clay, and carbonate bands (known as the marginal carbonates). However, there are regions where only clays are present, with only a weak to vanishing 2.5 μm band. This is marked "Clay only" on the Figure 7 correlation plot.
7. As seen in Figure 4 and 10, the IR spectrum for the clay stevensite (Tosca and Wright, 2018) is not consistent with that of Garde and Dourbes. Instead, as seen in Figure 10, we have shown that talc is more consistent with the clay signal in the VISIR spectra for Garde_207, and especially the Dourbes tailings from Sol 255.
8. Based on the lack of a 2.28 μm band, we can eliminate nontronite and hisingerite as the clays in Gardes and Dourbes.
9. Based on Table 2, showing the LIBS results for Cine/Garde and Ubraye/Dourbes, the low aluminum amounts for the olivine cumulate are not consistent with the presence of Al bearing clays, so we can rule out these smectite clay families:
    a.) saponite (Treiman et al., 2014)
    b.) beidellite,
    c.) stevensite (based on IR spectrum in Figure 4), and
    d.) montmorillonite (Bishop et al., 2008).
10. We have formulated a new conjecture that the olivine-clay-carbonate lithology is the missing earlier, more voluminous ultramafic rocks proposed by Rampey and Harvey. Needless to say, this will require further testing in situ.

## 6. Acknowledgements

We would like to thank Paolo Pilleri, Chip Legett and Dot DeLapp for their untiring work to keep the SuperCam tools and datasets up to date and ready for use in projects like this one. We would also like to thank François Poulet for helpful comments and suggestions on the manuscript. We also thank all of the people who contributed to the development and operations of the *Perseverance* rover. This work is supported by grants and contracts to the Mars 2020 project as part of the Mars Exploration Program.

## 7. References

Allwood, A.C., Wade, L.A., Foote, M.C., Elam, W.T., Hurowitz, J.A., Battel, S., Dawson, D.E., Denise, R.W., Ek, E.M., Gilbert, M.S., King, M.E., Liebe, C.C., Parker, T., Pedersen, D.A.K., Randall, D.P., Sharrow, R.F., Sondheim, M.E., Allen, G., Arnett, K., Au, M.H., Basset, C., Benn, M., Bousman, J.C., Braun, D., Calvet, R.J., Clark, B., Cinquini, L., Conaby, S., Conley, H.A., Davidoff, S., Delaney, J., Denver, T., Diaz, E., Doran, G.B., Ervin, J., Evans, M., Flannery, D.O., Gao, N., Gross, J., Grotzinger, J., Hannah, B., Harris, J.T., Harris, C.M., He, Y., Heirwegh, C.M., Hernandez, C., Hertzberg, E., Hodyss, R.P., Holden, J.R., Hummel, C., Jadusingh, M.A., Jørgensen, J.L., Kawamura, J.H., Kitiyakara, A., Kozaczek, K., Lambert, J.L., Lawson, P.R., Liu, Y., Luchik, T.S., Macneal, K.M., Madsen, S.N., McLennan, S.M., McNally, P., Meras, P.L., Muller, R.E., Napoli, J., Naylor, B.J., Nemere, P., Ponomarev, I., Perez, R.M., Pootrakul, N., Romero, R.A., Rosas, R., Sachs, J., Schaefer, R.T., Schein, M.E., Setterfield, T.P., Singh, V., Song, E., Soria, M.M., Stek, P.C., Tallarida, N.R., Thompson, D.R., Tice, M.M., Timmermann, L., Torossian, V., Treiman, A., Tsai, S., Uckert, K., Villalvazo, J., Wang, M., Wilson, D.W., Worel, S.C., Zamani, P., Zappe, M., Zhong, F., Zimmerman, R., 2020. PIXL: Planetary Instrument for X-Ray Lithochemistry. Space Sci. Rev. 216, 134. https://doi.org/10.1007/s11214-020-00767-7

Alwmark, S., 2022. Varied origins of Artuby signals complex series of geologic events in Jezero crater, Mars. J. Geophys. Res. this volume.

Amelin, Y., Krot, A.N., Hutcheon, I.D., Ulyanov, A.A., 2002. Lead Isotopic Ages of Chondrules and Calcium-Aluminum-Rich Inclusions. Science 297, 1678–1683. https://doi.org/10.1126/science.1073950

Anderson, R.B., Forni, O., Cousin, A., Wiens, R.C., Clegg, S.M., Frydenvang, J., Gabriel, T.S.J., Ollila, A., Schröder, S., Beyssac, O., Gibbons, E., Vogt, D.S., Clavé, E., Manrique, J.-A., Legett, C., Pilleri, P., Newell, R.T., Sarrao, J., Maurice, S., Arana, G., Benzerara, K., Bernardi, P., Bernard, S., Bousquet, B., Brown, A.J., Alvarez-Llamas, C., Chide, B., Cloutis, E., Comellas, J., Connell, S., Dehouck, E., Delapp, D.M., Essunfeld, A., Fabre, C., Fouchet, T., Garcia-Florentino, C., García-Gómez, L., Gasda, P., Gasnault, O., Hausrath, E.M., Lanza, N.L., Laserna, J., Lasue, J., Lopez, G., Madariaga, J.M., Mandon, L., Mangold, N., Meslin, P.-Y., Nelson, A.E., Newsom, H., Reyes-Newell, A.L., Robinson, S., Rull, F., Sharma, S., Simon, J.I., Sobron, P., Fernandez, I.T., Udry, A., Venhaus, D., McLennan, S.M., Morris, R.V., Ehlmann, B., 2022. Post-landing major element quantification using SuperCam laser induced breakdown spectroscopy. Spectrochim. Acta Part B At. Spectrosc. 188, 106347. https://doi.org/10.1016/j.sab.2021.106347

Arndt, N.T., Lesher, C.M., Houlé, M.G., Lewin, E., Lacaze, Y., 2004. Intrusion and crystallization of a spinifex-textured komatiite sill in Dundonald Township, Ontario. J. Petrol. 45, 2555–2571.

Bailey, S.W., 1988. Hydrous phyllosilicates:(exclusive of micas). Walter de Gruyter GmbH & Co

Bishop, J.L., Lane, M.D., Dyar, M.D., Brown, A.J., 2008. Reflectance and emission spectroscopy of four groups of phyllosilicates: smectites, kaolinite-serpentines, chlorites and micas. Clay Miner. 43, 35–54.

Bishop, J.L., Perry, K.A., Darby Dyar, M., Bristow, T.F., Blake, D.F., Brown, A.J., Peel, S.E., 2013. Coordinated spectral and XRD analyses of magnesite-nontronite-forsterite mixtures and implications for carbonates on Mars. J. Geophys. Res. Planets 635–650.

Borg, L.E., Connelly, J.N., Nyquist, L.E., Shih, C.-Y., Wiesmann, H., Reese, Y., 1999. The age of the carbonates in Martian meteorite ALH84001. Science 286, 90–94.

Bottinga, Y., Weill, D.F., 1972. The viscosity of magmatic silicate liquids: A model for calculation. Am. J. Sci. 272, 438–475.

Brindley, G.W., Bish, D.L., Wan, H.-M., 1977. The nature of kerolite, its relation to talc and stevensite. Mineral. Mag. 41, 443–452.

Brown, A. J., 2006. Spectral Curve Fitting for Automatic Hyperspectral Data Analysis. IEEE Trans. Geosci. Remote Sens. 44, 1601–1608. https://doi.org/10.1109/TGRS.2006.870435

Brown, A.J., 2006. Hyperspectral Mapping of Ancient Hydrothermal Systems (PhD). Macquarie University, Sydney, N.S.W.

Brown, A.J., Hook, S.J., Baldridge, A.M., Crowley, J.K., Bridges, N.T., Thomson, B.J., Marion, G.M., de Souza Filho, C.R., Bishop, J.L., 2010. Hydrothermal formation of Clay-Carbonate alteration assemblages in the Nili Fossae region of Mars. Earth Planet. Sci. Lett. 297, 174–182. https://doi.org/10.1016/j.epsl.2010.06.018

Brown, A.J., Viviano, C.E., Goudge, T.A., 2020. Olivine–Carbonate Mineralogy of the Jezero Crater Region. J. Geophys. Res. Planets 125, https://doi.org/10.1029/2019JE006011.

Brown, A.J., Walter, M.R., Cudahy, T.J., 2005. Hyperspectral Imaging Spectroscopy of a Mars Analog Environment at the North Pole Dome, Pilbara Craton, Western Australia. Aust. J. Earth Sci. 52, 353–364. https://doi.org/10.1080/08120090500134530

Calef, F.J., 2022. Crater retentions mechanisms in the Crater Floor - Fractured unit in Jezero Crater. J. Geophys. Res. this volume.

Citron, R.I., Manga, M., Hemingway, D.J., 2018. Timing of oceans on Mars from shoreline deformation. Nature 555, 643–646. https://doi.org/10.1038/nature26144

Clark, R.N., Swayze, G.A., Wise, R., Livo, E., Hoefen, T., Kokaly, R., Sutley, S.J., 2007. USGS digital spectral library splib06a: http://speclab.cr.usgs.gov/spectral.lib06., Digital Data Series 231. U.S. Geological Survey.

Clave, E., 2022. Carbonate detection with SuperCam in igneous rocks on the floor of Jezero Crater, Mars. J. Geophys. Res. this volume.

Corpolongo, A., 2022. SHERLOC Raman mineral detections of the Mars 2020 Crater Floor Campaign. J. Geophys. Res. this volume.

Crumpler, L.S., 2022. In Situ Geologic Context Mapping Transect on the Floor of Jezero Crater from Mars 2020/Perseverance Rover Observations. J. Geophys. Res. this volume.

Duke, E.F., 1994. Near infrared spectra of muscovite, Tschermak substitution, and metamorphic reaction progress: Implications for remote sensing. Geology 22, 621–624.

Ehlmann, B.L., Mustard, J.F., Murchie, S.L., Poulet, F., Bishop, J.L., Brown, A.J., Calvin, W.M., Clark, R.N., Marais, D.J.D., Milliken, R.E., Roach, L.H., Roush, T.L., Swayze, G.A., Wray, J.J., 2008. Orbital Identification of Carbonate-Bearing Rocks on Mars. Science 322, 1828–1832. https://doi.org/10.1126/science.1164759

Farley, K.A., Malespin, C., Mahaffy, P., Grotzinger, J.P., Vasconcelos, P.M., Milliken, R.E., Malin, M., Edgett, K.S., Pavlov, A.A., Hurowitz, J.A., Grant, J.A., Miller, H.B., Arvidson, R., Beegle, L., Calef, F., Conrad, P.G., Dietrich, W.E., Eigenbrode, J., Gellert, R., Gupta, S., Hamilton, V., Hassler, D.M., Lewis, K.W., McLennan, S.M., Ming, D., Navarro-Gonz√°lez, R., Schwenzer, S.P., Steele, A., Stolper, E.M., Sumner, D.Y., Vaniman, D., Vasavada, A., Williford, K., Wimmer-Schweingruber, R.F., the, M.S.L.S.T., 2013. In Situ Radiometric and Exposure Age Dating of the Martian Surface. Science. https://doi.org/10.1126/science.1247166

Farley, K.A., Williford, K.H., Stack, K.M., Bhartia, R., Chen, A., de la Torre, M., Hand, K., Goreva, Y., Herd, C.D.K., Hueso, R., Liu, Y., Maki, J.N., Martinez, G., Moeller, R.C., Nelessen, A., Newman, C.E., Nunes, D., Ponce, A., Spanovich, N., Willis, P.A., Beegle, L.W., Bell, J.F., Brown, A.J., Hamran, S.-E., Hurowitz, J.A., Maurice, S., Paige, D.A., Rodriguez-Manfredi, J.A., Schulte, M., Wiens, R.C., 2022. Aqueously altered igneous rocks on the floor of Jezero crater, Mars. Science in review.

Farley, K.A., Williford, K.H., Stack, K.M., Bhartia, R., Chen, A., de la Torre, M., Hand, K., Goreva, Y., Herd, C.D.K., Hueso, R., Liu, Y., Maki, J.N., Martinez, G., Moeller, R.C., Nelessen, A., Newman, C.E., Nunes, D., Ponce, A., Spanovich, N., Willis, P.A., Beegle, L.W., Bell, J.F., Brown, A.J., Hamran, S.-E., Hurowitz, J.A., Maurice, S., Paige, D.A., Rodriguez-Manfredi, J.A., Schulte, M., Wiens, R.C., 2020. Mars 2020 Mission Overview. Space Sci. Rev. 216, 142. https://doi.org/10.1007/s11214-020-00762-y

García-Romero, E., Lorenzo, A., García-Vicente, A., Morales, J., García-Rivas, J., Suárez, M., 2021. On the structural formula of smectites: a review and new data on the influence of exchangeable cations. J. Appl. Crystallogr. 54, 251–262. https://doi.org/10.1107/S1600576720016040

Goudge, T.A., Mohrig, D., Cardenas, B.T., Hughes, C.M., Fassett, C.I., 2018. Stratigraphy and paleohydrology of delta channel deposits, Jezero crater, Mars. Icarus 301, 58–75.

Goudge, T.A., Mustard, J.F., Head, J.W., Fassett, C.I., Wiseman, S.M., 2015. Assessing the Mineralogy of the Watershed and Fan Deposits of the Jezero Crater Paleolake System, Mars. J. Geophys. Res. Planets 2014JE004782. https://doi.org/10.1002/2014JE004782

Hamilton, V.E., Christensen, P.R., 2005. Evidence for extensive, olivine-rich bedrock on Mars. Geology 33, 433–436.

Hiesinger, H., Head, J.W., 2004. The Syrtis Major volcanic province, Mars: Synthesis from Mars Global Surveyor data. J. Geophys. Res. 109, 10.1029/2003JE002143.

Hoefen, T.M., Clark, R.N., Bandfield, J.L., Smith, M.D., Pearl, J.C., Christensen, P.R., 2003. Discovery of Olivine in the Nili Fossae Region of Mars. Science 302, 627–630.

Horgan, B.H.N., Anderson, R.B., Dromart, G., Amador, E.S., Rice, M.S., 2020. The mineral diversity of Jezero crater: Evidence for possible lacustrine carbonates on Mars. Icarus 113526. https://doi.org/10.1016/j.icarus.2019.113526

Huppert, H.E., Sparks, R.S.J., 1985. Cooling and contamination of mafic and ultramafic magmas during ascent through continental crust. Earth Planet. Sci. Lett. 74, 371–386.

Huppert, H.E., Sparks, R.S.J., 1981. The fluid dynamics of a basaltic magma chamber replenished by influx of hot, dense ultrabasic magma. Contrib. Mineral. Petrol. 75, 279–289. https://doi.org/10.1007/BF01166768

Huppert, H.E., Sparks, R.S.J., Turner, J.S., Arndt, N.T., 1984. Emplacement and cooling of komatiite lavas. Nature 309, 13–16.

Isherwood, R.J., Jozwiak, L.M., Jansen, J.C., Andrews-Hanna, J.C., 2013. The volcanic history of Olympus Mons from paleo-topography and flexural modeling. Earth Planet. Sci. Lett. 363, 88–96.

Kah, L.C., 2022. Use of size-frequency distributions in the interpretation of planetary surface materials. J. Geophys. Res. this volume.

Kremer, C.H., Mustard, J.F., Bramble, M.S., 2019. A widespread olivine-rich ash deposit on Mars. Geology. https://doi.org/10.1130/G45563.1

Lapen, T.J., Righter, M., Brandon, A.D., Debaille, V., Beard, B.L., Shafer, J.T., Peslier, A.H., 2010. A Younger Age for ALH84001 and Its Geochemical Link to Shergottite Sources in Mars. Science 328, 347–351.

Liu, Y., 2022. An olivine cumulate outcrop on the floor of Jezero crater, Mars. Science in review.

Mandon, L., 2022. Reflectance of Jezero crater floor: 2. Mineralogical interpretation. J. Geophys. Res. this volume.

Mandon, L., Beck, P., Quantin-Nataf, C., Dehouck, E., Thollot, P., Loizeau, D., Volat, M., 2022. ROMA: A Database of Rock Reflectance Spectra for Martian In Situ Exploration. Earth Space Sci. 9, e2021EA001871. https://doi.org/10.1029/2021EA001871

Mandon, L., Quantin-Nataf, C., Thollot, P., Mangold, N., Lozac'h, L., Dromart, G., Beck, P., Dehouck, E., Breton, S., Millot, C., Volat, M., 2020. Refining the age, emplacement and alteration scenarios of the olivine-rich unit in the Nili Fossae region, Mars. Icarus 336, 113436. https://doi.org/10.1016/j.icarus.2019.113436

Mangold, N., Gupta, S., Gasnault, O., Dromart, G., Tarnas, J.D., Sholes, S.F., Horgan, B., Quantin-Nataf, C., Brown, A.J., Le Mouélic, S., 2021. Perseverance rover reveals an ancient delta-lake system and flood deposits at Jezero crater, Mars. Science eabl4051.

Mangold, N., Poulet, F., Mustard, J.F., Bibring, J.P., Gondet, B., Langevin, Y., Ansan, V., Masson, P., Fassett, C., Head, J.W., III, Hoffmann, H., Neukum, G., 2007. Mineralogy of the Nili Fossae region with OMEGA/Mars Express data: 2. Aqueous alteration of the crust. J. Geophys. Res. 112, doi:10.1029/2006JE002835.

Martin, P.E., Farley, K.A., Baker, M.B., Malespin, C.A., Schwenzer, S.P., Cohen, B.A., Mahaffy, P.R., McAdam, A.C., Ming, D.W., Vasconcelos, P.M., Navarro-González, R., 2017. A Two-Step K-Ar Experiment on Mars: Dating the Diagenetic Formation of Jarosite from Amazonian Groundwaters. J. Geophys. Res. Planets 122, 2803–2818. https://doi.org/10.1002/2017JE005445

Maurice, S., Wiens, R.C., Bernardi, P., Caïs, P., Robinson, S., Nelson, T., Gasnault, O., Reess, J.-M., Deleuze, M., Rull, F., 2021. The SuperCam instrument suite on the Mars 2020 rover: Science objectives and Mast-Unit description. Space Sci. Rev. 217, 1–108.

Mustard, J.F., Poulet, F., Head, J.W., Mangold, N., Bibring, J.P., Pelkey, S.M., Fassett, C.I., Langevin, Y., Neukum, G., 2007. Mineralogy of the Nili Fossae region with OMEGA/Mars Express data: 1. Ancient impact melt in the Isidis Basin and implications for the transition from the Noachian to Hesperian. J. Geophys. Res. 112.

Núñez, J., 2022. Stratigraphy and Mineralogy of the Séítah formation on the floor of Jezero crater, Mars, as seen with Mastcam-Z. J. Geophys. Res. this volume.

Nyquist, L.E., Shih, C.-Y., McCubbin, F.M., Santos, A.R., Shearer, C.K., Peng, Z.X., Burger, P.V., Agee, C.B., 2016. Rb-Sr and Sm-Nd isotopic and REE studies of igneous components in the bulk matrix domain of Martian breccia Northwest Africa 7034. Meteorit. Planet. Sci. 51, 483–498. https://doi.org/10.1111/maps.12606

Pinet, P., Chevrel, S., 1990. Spectral Identification of Geological Units on the Surface of Mars Related to the Presence of Silicates From Earth-Based Near-Infrared Telescopic Charge-Coupled Device Imaging. J. Geophys. Res. 95, 14435.

Pinkerton, H., Wilson, L., 1994. Factors controlling the lengths of channel-fed lava flows. Bull. Volcanol. 56, 108–120. https://doi.org/10.1007/BF00304106

Plescia, J.B., 1990. Recent flood lavas in the Elysium region of Mars. Icarus 88, 465–490.

Potin, S., Brissaud, O., Beck, P., Schmitt, B., Magnard, Y., Correia, J.-J., Rabou, P., Jocou, L., 2018. SHADOWS: a spectro-gonio radiometer for bidirectional reflectance studies of dark meteorites and terrestrial analogs: design, calibrations, and performances on challenging surfaces. Appl. Opt. 57, 8279–8296.

Putirka, K., 2016. Rates and styles of planetary cooling on Earth, Moon, Mars, and Vesta, using new models for oxygen fugacity, ferric-ferrous ratios, olivine-liquid Fe-Mg exchange, and mantle potential temperature. Am. Mineral. 101, 819–840. https://doi.org/10.2138/am-2016-5402

Rampey, M., Harvey, R., 2012. Mars Hesperian Magmatism as Revealed by Syrtis Major and the Circum-Hellas Volcanic Province. Earth Moon Planets 109, 61–75. https://doi.org/10.1007/s11038-012-9404-0

Ravanis, E., Fagents, S.A., Newman, C.E., Horgan, B.H., Holm-Alwmark, S., Brown, A.J., Rice Jr, J.W., Mandon, L., Zorzano, M.-P., 2022. The Potential for Pyroclastic Deposits in the Jezero Crater Region of Mars from Ash Dispersal Modeling, in: LPSC. Presented at the 53rd Lunar and Planetary Science Conference, LPI, Houston, TX, p. Abstract #1692.

Robbins, S.J., Achille, G.D., Hynek, B.M., 2011. The volcanic history of Mars: High-resolution crater-based studies of the calderas of 20 volcanoes. Icarus 211, 1179–1203. https://doi.org/10.1016/j.icarus.2010.11.012

Rogers, D., Warner Nicholas H., Golombek Matthew P., Head James W., Cowart Justin C., 2018. Areally Extensive Surface Bedrock Exposures on Mars: Many Are Clastic Rocks, Not Lavas. Geophys. Res. Lett. 45, 1767–1777. https://doi.org/10.1002/2018GL077030

Roush, T.L., Bishop, J.L., Brown, A.J., Blake, D.F., Bristow, T.F., 2015. Laboratory reflectance spectra of clay minerals mixed with Mars analog materials: Toward enabling quantitative clay abundances from Mars spectra. Icarus 258, 454–466. https://doi.org/10.1016/j.icarus.2015.06.035

Royer, C., 2022. Reflectance of Jezero crater floor: 1. Data processing and calibration of IRS/SuperCam. J. Geophys. Res. this volume.

Ruff, S.W., Hamilton, V.E., Rogers, A.D., Edwards, C.S., Horgan, B.H.N., 2022. Olivine and carbonate-rich bedrock in Gusev crater and the Nili Fossae region of Mars may be altered ignimbrite deposits. Icarus 380, 114974. https://doi.org/10.1016/j.icarus.2022.114974

Shahrzad, S., Kinch, K.M., Goudge, T.A., Fassett, C.I., Needham, D.H., Quantin–Nataf, C., Knudsen, C.P., 2019. Crater statistics on the dark-toned, mafic floor unit in Jezero Crater, Mars. Geophys. Res. Lett. 0. https://doi.org/10.1029/2018GL081402

Sholes, S.F., Dickeson, Z.I., Montgomery, D.R., Catling, D.C., 2021. Where are Mars' hypothesized ocean shorelines? Large lateral and topographic offsets between different versions of paleoshoreline maps. J. Geophys. Res. Planets 126, e2020JE006486.

Sholes, S.F., Rivera-Hernández, F., 2022. Constraints on the uncertainty, timing, and magnitude of potential Mars oceans from topographic deformation models. Icarus 378, 114934.

Simon, J.I., 2022. Samples Collected from the Floor of Jezero Crater with the Mars 2020 Perseverance Rover. J. Geophys. Res. this volume.

Stack, K.M., Williams, N.R., Calef, F., Sun, V.Z., Williford, K.H., Farley, K.A., Eide, S., Flannery, D., Hughes, C., Jacob, S.R., Kah, L.C., Meyen, F., Molina, A., Nataf, C.Q., Rice, M., Russell, P., Scheller, E., Seeger, C.H., Abbey, W.J., Adler, J.B., Amundsen, H., Anderson, R.B., Angel, S.M., Arana, G., Atkins, J., Barrington, M., Berger, T., Borden, R., Boring, B., Brown, A., Carrier, B.L., Conrad, P., Dypvik, H., Fagents, S.A., Gallegos, Z.E., Garczynski, B., Golder, K., Gomez, F., Goreva, Y., Gupta, S., Hamran, S.-E., Hicks, T., Hinterman, E.D., Horgan, B.N., Hurowitz, J., Johnson, J.R., Lasue, J., Kronyak, R.E., Liu, Y., Madariaga, J.M., Mangold, N., McClean, J., Miklusicak, N., Nunes, D., Rojas, C., Runyon, K., Schmitz, N., Scudder, N., Shaver, E., SooHoo, J., Spaulding, R., Stanish, E., Tamppari, L.K., Tice, M.M., Turenne, N., Willis, P.A., Aileen Yingst, R., 2020.

Photogeologic Map of the Perseverance Rover Field Site in Jezero Crater Constructed by the Mars 2020 Science Team. Space Sci. Rev. 216, 127. https://doi.org/10.1007/s11214-020-00739-x

Sun, V.Z., 2022. Exploring the Jezero Crater Floor: Summary and Synthesis of Perseverance's First Science Campaign. J. Geophys. Res. this volume.

Tarnas, J.D., Stack, K.M., Parente, M., Koeppel, A.H.D., Mustard, J.F., Moore, K.R., Horgan, B.H.N., Seelos, F.P., Cloutis, E.A., Kelemen, P.B., Flannery, D., Brown, A.J., Frizzell, K.R., Pinet, P., 2021. Characteristics, Origins, and Biosignature Preservation Potential of Carbonate-Bearing Rocks Within and Outside of Jezero Crater. J. Geophys. Res. Planets 126, e2021JE006898. https://doi.org/10.1029/2021JE006898

Tice, M.M., 2022. Primary and Alteration Textures of Séítah Formation Rocks Inferred by X-ray Fluorescence and Diffraction. Sci. Adv. in review.

Tornabene, L.L., Moersch, J.E., McSween, H.Y., Hamilton, V.E., Piatek, J.L., Christensen, P.R., 2008. Surface and crater-exposed lithologic units of the Isidis Basin as mapped by coanalysis of THEMIS and TES derived data products. J. Geophys. Res. Planets 113, E10001. https://doi.org/10.1029/2007JE002988

Tosca, N., 2022. Serpentinization of the ancient Jezero Crater floor. J. Geophys. Res. this volume.

Tosca, N.J., Wright, V.P., 2018. Diagenetic pathways linked to labile Mg-clays in lacustrine carbonate reservoirs: a model for the origin of secondary porosity in the Cretaceous pre-salt Barra Velha Formation, offshore Brazil. Geol. Soc. Lond. Spec. Publ. 435, 33–46. https://doi.org/10.1144/SP435.1

Treiman, A.H., 2005. The nakhlite meteorites: Augite-rich igneous rocks from Mars. Geochemistry 65, 203–270. https://doi.org/10.1016/j.chemer.2005.01.004

Treiman, A.H., Morris, R.V., Agresti, D.G., Graff, T.G., Achilles, C.N., Rampe, E.B., Bristow, T.F., Ming, D.W., Blake, D.F., Vaniman, D.T., Bish, D.L., Chipera, S.J., Morrison, S.M., Downs, R.T., 2014. Ferrian saponite from the Santa Monica Mountains (California, U.S.A., Earth): Characterization as an analog for clay minerals on Mars with application to Yellowknife Bay in Gale Crater. Am. Mineral. 99, 2234–2250. https://doi.org/10.2138/am-2014-4763

Turrenne, N., Sidhu, S., Applin, D.M., Cloutis, E.A., Mertzman, S.A., Hausrath, E.M., 2022. Spectral properties of Nontronite under Earth surface conditions, Mars-like surface conditions, and (impact) heating events: Implications for instrument detections on Mars. Icarus in prep.

Udry, A., 2022. A Mars2020 Perseverance SuperCam Perspective on the Igneous Nature of the Maaz formation at Jezero crater and link with Seitah, Mars. J. Geophys. Res. this volume.

Werner, S.C., 2008. The early martian evolution—Constraints from basin formation ages. Icarus 195, 45–60. https://doi.org/10.1016/j.icarus.2007.12.008

Wichman, R.W., Schultz, P.H., 1989. Sequence and mechanisms of deformation around the Hellas and Isidis Impact Basins on Mars. J. Geophys. Res. Solid Earth 94, 17333–17357. https://doi.org/10.1029/JB094iB12p17333

Wiens, R.C., 2022. Compositionally Stratified terrain in Jezero Crater. Sci. Adv. Accepted.

Wiens, R.C., Maurice, S., Robinson, S.H., Nelson, A.E., Cais, P., Bernardi, P., Newell, R.T., Clegg, S., Sharma, S.K., Storms, S., 2021. The SuperCam instrument suite on the NASA Mars 2020 rover: Body unit and combined system tests. Space Sci. Rev. 217, 1–87.

Williford, K.H., Farley, K.A., Stack, K.M., Allwood, A.C., Beaty, D., Beegle, L.W., Bhartia, R., Brown, A.J., de la Torre Juarez, M., Hamran, S.-E., Hecht, M.H., Hurowitz, J.A., Rodriguez-Manfredi, J.A., Maurice, S., Milkovich, S., Wiens, R.C., 2018. Chapter 11 - The NASA Mars 2020 Rover Mission and the Search for Extraterrestrial Life, in: Cabrol, N.A., Grin, E.A. (Eds.), From Habitability to Life on Mars. Elsevier, pp. 275–308. https://doi.org/10.1016/B978-0-12-809935-3.00010-4

Wilson, L., Head, J.W., 1994. Mars - Review and Analysis of Volcanic-Eruption Theory and Relationships to Observed Landforms. Rev. Geophys. 32, 221–263.

Worster, M.G., Huppert, H.E., Sparks, R.S.J., 1993. The crystallization of lava lakes. J. Geophys. Res. Solid Earth 98, 15891–15901. https://doi.org/10.1029/93JB01428

Zastrow, A.M., Glotch, T.D., 2021. Distinct carbonate lithologies in Jezero crater, Mars. Geophys. Res. Lett. 48, e2020GL092365.